\documentclass{jfm}

\usepackage{graphicx}
\usepackage{natbib}
\usepackage{color}
\usepackage{pdflscape}

\usepackage{amsmath}
\usepackage{amssymb}
\usepackage{afterpage}

\newcommand{\be}{\begin{equation}}
\newcommand{\ee}{\end{equation}}
\newcommand{\bs}{\begin{subequations}}
\newcommand{\es}{\end{subequations}}

\newcommand{\rmd}{\mathrm{d}}
\newcommand{\rme}{\mathrm{e}}
\newcommand{\rmi}{\mathrm{i}}

\newcommand{\half}{{\textstyle\frac12}}
\newcommand{\quarter}{{\textstyle\frac14}}

\newcommand{\bcdot}{{\boldsymbol\cdot}}
\newcommand{\btimes}{{\boldsymbol\times}}
\newcommand{\bk}{{\boldsymbol k}}
\newcommand{\bV}{{\boldsymbol V}}
\newcommand{\bK}{{\boldsymbol K}}
\newcommand{\bX}{{\boldsymbol X}}
\newcommand{\bv}{{\boldsymbol v}}
\newcommand{\bx}{{\boldsymbol x}}
\newcommand{\bxi}{{\boldsymbol \xi}}

\newcommand{\efac}{\rme^{\rmi\bk\bcdot{\boldsymbol \xi}}}
\newcommand{\Fr}{\mathrm{Fr}}
\newcommand{\Frs}{\mathrm{Fr}_\mathrm{s}}
\newcommand{\Frh}{\mathrm{Fr}_h}

\newcommand{\Frsb}{\mathrm{Fr}_{Sb}}
\newcommand{\FK}{\Phi_\mathrm{K}}

\newcommand{\pipiint}{\int_{-\upi}^{\upi}}
\newcommand{\Kint}{\int_0^\infty \rmd K}
\newcommand{\kint}{\int\frac{\rmd^2 k}{(2\upi)^2}}

\newcommand{\hu}{\hat{u}}
\newcommand{\hv}{\hat{v}}
\newcommand{\hw}{\hat{w}}
\newcommand{\hp}{\hat{p}}
\newcommand{\hpext}{\hat{p}_\text{ext}}
\newcommand{\pext}{p_\text{ext}}

\newcommand{\Vc}{V_\mathrm{crit}}

\newcommand{\kV}{\bk\bcdot\bV}
\newcommand{\kUpar}{(k_x U-\kV)}

\newcommand{\Isd}{I_\text{s.d.}}
\newcommand{\Ic}{I_\text{conn.}}
\newcommand{\Ksd}{K_\text{s.d.}}
\newcommand{\zff}{\zeta_\mathrm{f.f.}}

\DeclareMathOperator{\sech}{sech}
\DeclareMathOperator{\sg}{Sg}
\DeclareMathOperator*{\Res}{Res}
\newcommand*{\res}{\Res\limits}

\ifCUPmtlplainloaded \else
  \checkfont{eurm10}
  \iffontfound
    \IfFileExists{upmath.sty}
      {\typeout{^^JFound AMS Euler Roman fonts on the system,
                   using the 'upmath' package.^^J}%
       \usepackage{upmath}}
      {\typeout{^^JFound AMS Euler Roman fonts on the system, but you
                   dont seem to have the}%
       \typeout{'upmath' package installed. JFM.cls can take advantage
                 of these fonts,^^Jif you use 'upmath' package.^^J}%
       \providecommand\upi{\upi}%
      }
  \else
    \providecommand\upi{\upi}%
  \fi
\fi

\ifCUPmtlplainloaded \else
  \checkfont{msam10}
  \iffontfound
    \IfFileExists{amssymb.sty}
      {\typeout{^^JFound AMS Symbol fonts on the system, using the
                'amssymb' package.^^J}%
       \usepackage{amssymb}%
         \let\leq=\leqslant
         \let\geq=\geqslant
      }{}
  \fi
\fi

\ifCUPmtlplainloaded \else
  \IfFileExists{amsbsy.sty}
    {\typeout{^^JFound the 'amsbsy' package on the system, using it.^^J}%
     \usepackage{amsbsy}}
    {\providecommand\boldsymbol[1]{\mbox{\boldmath $##1$}}}
\fi

\title[Ship waves on uniform shear current]{Ship waves on uniform shear current at finite depth: wave resistance and critical velocity}
\author[Y. Li and S. \AA. Ellingsen]{Yan Li$^1$ and Simen \AA. Ellingsen$^1$\thanks{Email address for correspondence: simen.a.ellingsen@ntnu.no}}

\affiliation{$^1$Department of Energy and Process Engineering, Norwegian University of Science and Technology, N-7491 Trondheim, Norway}

\begin{document}

\maketitle
\begin{abstract}
  We present a comprehensive theory for linear gravity-driven ship waves in the presence of a shear current with uniform vorticity, including the effects of finite water depth. The wave resistance in the presence of shear current is calculated for the first time, containing in general a non-zero lateral component. While formally apparently a straightforward extension of existing deep water theory, the introduction of finite water depth is physically non-trivial, since the surface waves are now affected by a subtle interplay of the effects of the current and the sea bed. This becomes particularly pronounced when considering the phenomenon of critical velocity, the velocity at which transversely propagating waves become unable to keep up with the moving source. The phenomenon is well known for shallow water, and was recently shown to exist also in deep water in the presence of a shear current [Ellingsen, J.~Fluid Mech.\ {\bf 742} R2 (2014)]. We derive the exact criterion for criticality as a function of an intrinsic shear Froude number $S\sqrt{b/g}$ ($S$ is uniform vorticity, $b$ size of source), the water depth, and the angle between the shear current and the ship's motion. 
  
  Formulae for both the normal and lateral wave resistance force are derived, and we analyse its dependence on the source velocity (or Froude number $\Fr$) for different amounts of shear and different directions of motion. 
  The effect of the shear current is to increase wave resistance for upstream ship motion and decrease it for downstream motion. Also the value of $\Fr$ at which $R$ is maximal is lowered for upstream and increased for downstream directions of ship motion. For oblique angles between ship motion and current there is a lateral wave resistance component which can amount to $10$-$20\%$ of the normal wave resistance for side-on shear and $S\sqrt{b/g}$ of order unity.
  
  The theory is fully laid out and far-field contributions are carefully separated off by means of Cauchy's integral theorem, exposing potential pitfalls associated with a slightly different method (Sokhotsky--Plemelj) used in several previous works.
\end{abstract}

\section{Introduction}

Recent times have seen a resurgence of interest in ship waves, the phenomenon whose theory was pioneered by Lord Kelvin well over a century ago \citep{kelvin1887}. A topic of particular interest recently has been the angle formed by a ship's waves. Famously, in deep waters the gravity-driven waves behind a ship were shown by Kelvin to lie within a sector of half angle $\FK=19^\circ28'$, regardless of the ship's size, shape and velocity. \citet{rabaud13} remarked, however, that images of ship wakes indicate the wake angle narrowing with increasing Froude number $\Fr=V/\sqrt{gb}$, $V$ being the ship's velocity and $b$ its size. The issue was soon resolved by \citet{darmon14} and \citet{noblesse14} who demonstrated that while Kelvin's result remains true, the angle at which the waves' amplitude is greatest is smaller than Kelvin's angle, and decreases with increasing $\Fr$, making the wake appear narrower. The phenomenon was in fact observed and analysed already several decades ago \citep{munk87,brown89,reed02}, and number of authors have further elucidated this question recently \citep[e.g.,][]{moisy14,benzaquen14,pethiyagoda14,he15,zhu15,pethiyagoda15}. In particular it has been shown that interference effects between waves generated at the bow and the stern determine the scaling of the apparent angle with $\Fr$ \citep{noblesse14,zhang15,zhu15}.

Although a large literature exists on waves on shear currents in two dimensions or when the shear is horizontal \citep[c.f., e.g.,][and references therein]{peregrine76,buhler09,ellingsen14c}, previous knowledge of waves on vertically sheared currents in three dimensions is very scarce and limited to a few scattered references \citep[e.g.,][]{craik68,johnson90,mchugh94}. However, it was recently demonstrated that, when viscosity is neglected, a general solution to linear wave problems exists when a shear current with uniform vorticity is present, and  the solution was used to solve the problems of ship waves \citep{ellingsen14a}, oscillating point source \citep{ellingsen15b}, and initial value problems \citep{ellingsen14b,li15}.  The presence of a shear current was found to influence the waves behind a ship profoundly. A ship travelling against the shear current (seen from a coordinate system where the unperturbed water surface is at rest) produces longer transverse wavelengths, and its wake is broader than Kelvin's angle $\FK$, and \emph{vice versa} for a ship travelling with the current. When the shear current makes an angle $\beta$ other than $0$ or $\upi$ with the ship's line of motion, the wake is found to be asymmetrical, its angular extent being greater than $\FK$ on one side and smaller on the other. Moreover it was shown that, except at $\beta=\upi$, there exists a critical velocity at which the Kelvin wake angle reaches a total angle of $180^\circ$, beyond which the ship moves too fast for the transverse part of the ship waves to keep up, thus being unable to contribute to a stationary wake. 

\begin{figure}
  \centerline{\includegraphics[width=3.5in]{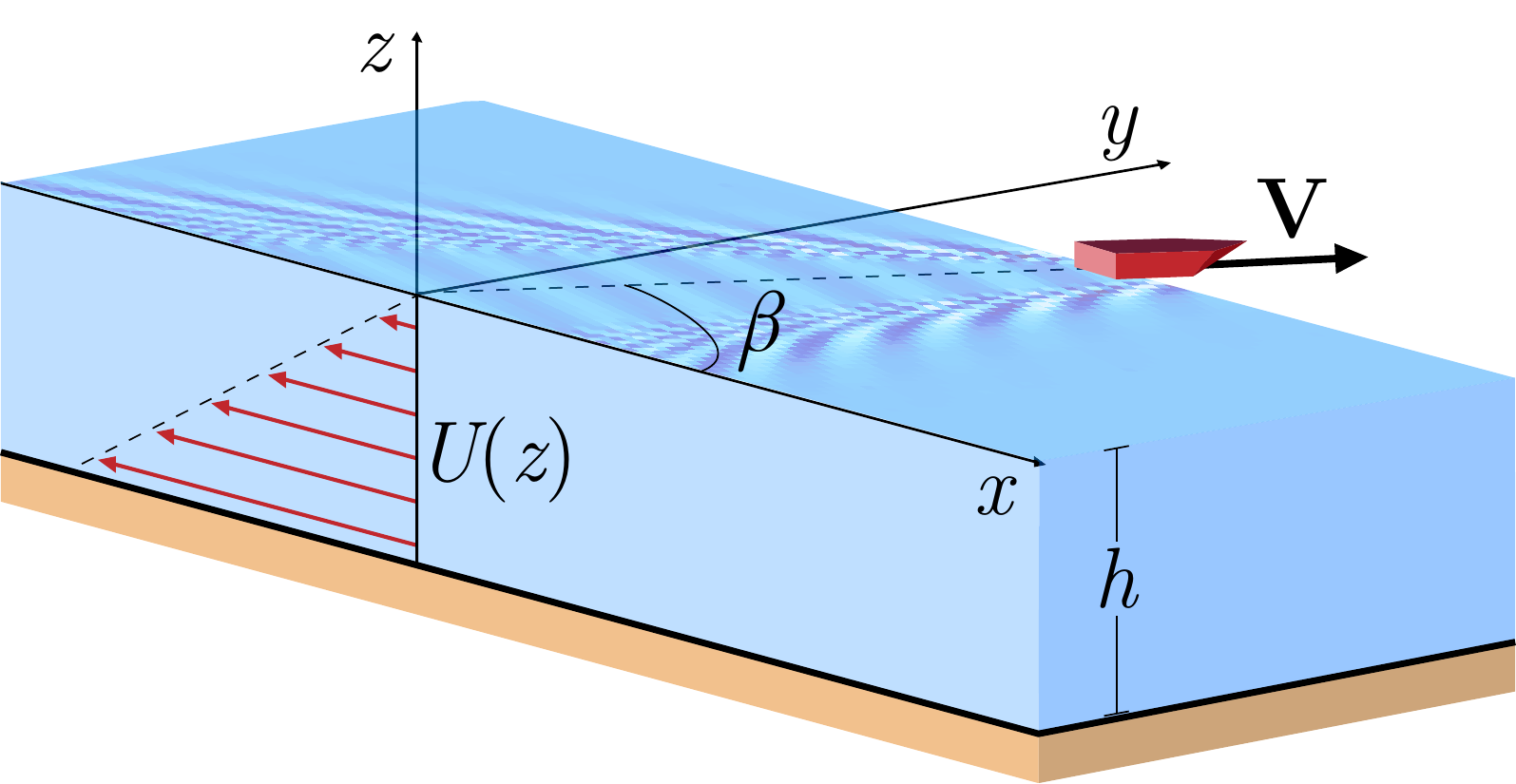}}
  \caption{The geometry considered: a boat (modelled as a pressure disturbance) travelling at velocity $\bV$ making an angle $\beta$ with an underlying shear current of uniform vorticity. The undisturbed surface is chosen to be at rest with respect to the coordinate system.}
  \label{fig:geom}
\end{figure}

Waves carry momentum, so a moving wave source must feel a resistance force equal to the rate at which impulse is imparted to the waves which are created. This wave resistance force typically accounts for more than 30\% of the fuel consumption of large sea-going vessels \citep{faltinsen05}. Knowing that the train of waves is affected by the presence of a sub-surface shear current, it seems likely that also the wave resistance will be affected by the current, a notion which we confirm and quantify herein. In particular, when the ship's line of motion is not parallel with the current the ship waves are asymmetrical, and the wave resistance also obtains a lateral component which, our calculations indicate, can amount to a significant percentage of the normal resistance force. 

We present in the current paper a reasonably complete theory of linear ship waves in the presence of uniform vorticity when also the water is assumed to have a finite, constant depth $h$. The geometry is shown in Fig.~\ref{fig:geom}. While accounting for the finite depth is a straightforward extension of the formal derivation of general results, the physical implications are highly non-trivial, and introduce a subtle interplay between the effects of the shear current and the sea bed upon the surface elevation. We begin by laying out the theory in its general form and pay particular attention to the separation into near-field and far-field by careful use of the Cauchy integral theorem. The resulting far-field expression for the surface elevation differs slightly from that obtained by an alternative procedure (the Sokhotsky--Plemelj formula) used repeatedly in the literature, and we explain why the latter procedure is in fact potentially treacherous when employed, as here, in the context of the radiation condition. 

We thereupon discuss in detail the phenomenon of critical velocity, which can occur whenever the dispersion relation makes the phase velocity bounded for all wavelengths in at least certain directions. Ship waves are termed supercritical when, for wave components in a finite sector of propagation directions, the ship's velocity is greater than the maximum phase velocity measured along the direction of motion, in which case transverse waves will vanish from the ship wake completely. We derive explicit conditions for criticality and the critical velocity as a function of vorticity $S$, depth and the angle $\beta$. 

Numerical evaluations of ship waves are thereafter carried out with particular emphasis on the transition between critical and non-critical waves; when the water depth is finite, increasing the shear can result in a transision both from sub-critical to supercritical (as reported by \citet{ellingsen14a}), or in certain cases also from supercritical to sub-critical. We finally calculate the wave resistance on the model ``ship'', both the standard resistance force to the forward motion and the lateral force resulting from asymmetric wave-making. While a realistic wave resistance calculation for a real vessel must take account of the actual hull shape, which is beyond our present scope, the calculations show trends which are likely to hold in general. Firstly, that wave resistance is increased when the ship motion has an upstream component (as seen from the system where the unisturbed surface is at rest), and decreased for downstream ship motion. Secondly, the Froude number at which the wave resistance is maximal is lowered for upstream and increased for downstream ship velocity.

\section{Mathematical model and general solution}

This section lays out the mathematical theory. The mathematical model and its general solution for wave pattern and wave resistance are derived in Sections \ref{sec:general} and \ref{sec:WRgen} with Eqs.~\eqref{zetaGen}, \eqref{Rgen} and \eqref{Rpgen} as final results. A discussion of the dispersion relation ensues in Section \ref{sec:disprel}, whereupon far-field expressions are derived in Section \ref{sec:nearfar}. It was seen as necessary to recount the careful extraction of the far-field waves in some detail herein, in order to highlight pitfalls and rectify errors associated with cavalier use of a related method used in recent literature. The final expression, Eq.~\eqref{zetafar2}, is found only after also taking on board lessons from Section \ref{sec:crit}.

\subsection{General solution}\label{sec:general}

For fully three-dimensional flow in the presence of vorticity, potential theory is not an option, so to solve the flow problem we must turn to the Euler equations, describing inviscid flow. 
The flow is assumed to be incompressible. We write the full velocity and pressure field on the form
\be
  \bv = (U(z) + \hu, \hv,\hw); ~~~ P = -\rho gz + \hp; ~~~ U(z) = Sz.
\ee
Here $U(z)$ is the basic shear current of constant vorticity $S\geq 0$, and the hatted quantities are perturbations due to the waves. We shall assume all perturbations to be small, and work to linear order in these quantities. We have assumed the surface of the water to be at rest with respect to the coordinate system in order that results can be immediately compared to previous work: this is easily generalised by an overall Galilean coordinate transformation.

The wave source (``ship'') is modelled as a superimposed localised pressure $\hpext$ of constant shape and strength, travelling with velocity $\bV$ which makes an angle $\beta$ with the $x$ axis, and hence the shear flow. The situation is sketched in figure \ref{fig:geom}. We consider only stationary solutions as seen from the boat, hence all physical quantities will depend on surface position $\bx=(x,y)$ and time $t$ only through the combination $\bxi = \bx-\bV t$. 
Such a stationary wake may be interpreted as a continuous series of ring waves emitted by the travelling source, and all our results might instead be derived based on such a formalism \cite{li15b}.

The flow is a solution to the Euler equation
\be
  \partial_t\bv + (\bv\bcdot\nabla)\bv = -\nabla (P/\rho + gz)
\ee
which we linearize with respect to perturbations.
We use a Fourier decomposition of perturbation quantities into plane waves according to
\be\label{fourier}
  [\hu,\hv,\hw,\hp](\bxi,z) =\kint [u,v,w,p](\bk,z)\efac.
\ee
Following \citet{ellingsen14a}, we can eliminate $u, v$ and $p$ to find a simple Rayleigh equation (the inviscid form of the Orr-Sommerfeldt equation) for $w$ alone, $ w'' = k^2 w.$
Solving this subject to the boundary condition that $w(\bk,-h)=0$ (no vertical velocity at the bottom), we find the solution to the full flow field as follows
\bs\label{Euler}
\begin{align}
  u(\bk,z) =& \rmi A(\bk)\left[k_x\cosh k(z+h) + \frac{S k_y^2\sinh k(z+h)}{k\kUpar} \right],\\
  v(\bk,z) =& \rmi A(\bk)\left[k_y\cosh k(z+h) - \frac{S k_xk_y\sinh k(z+h)}{k\kUpar} \right],\\
  w(\bk,z) =& k A(\bk)\sinh k(z+h) ,\\
  p(\bk,z) =&-\rmi A(\bk)\left[\kUpar \cosh k(z+h) - \frac{Sk_x}{k}\sinh k(z+h)\right].
\end{align}
\es
Here, $A(\bk)$ is an unknown coefficient. These solutions are the ship wave equivalents of the general solutions reported by \citet{ellingsen14b}. Note that the motion introduced by this wave solution is itself rotational since it shifts and twists the vortex lines of the background flow, unlike any wave motion described by potential theory \citep{ellingsen16}.

Let the surface elevation (relative to its equilibrium state) be $\zeta(\bxi)$ and the external pressure be $\hpext(\bxi)$, and let their Fourier transforms in the manner of Eq.~\eqref{fourier}, be $B(\bk)$ and $\pext(\bk)$, respectively.
We can now write down the linearised kinematic boundary condition (stating that a particle on the surface stays on the surface)
\be
  k A(\bk)\sinh kh = -\rmi (\kV) B(\bk)
\ee
(note that $U(0)=0$ by choice),
and dynamic boundary condition (stating that normal stress, as given by the pressure through Euler equation, is continuous at the surface),
\be
  \rmi A(\bk)\left[\kV\cosh kh + \frac{S k_x}{k}\sinh kh\right] - g B(\bk) = \pext(\bk)/\rho.
\ee
Eliminating $A(\bk)$, we find $B(\bk)$ which we integrate over the $\bk$ plane to find $\zeta$. One now encounters the same difficulty always encountered when considering waves in quasi-stationary or quasi-periodiodic wave systems, namely that the integral is indeterminate due to poles on the axes of integration. The criterion that our system, while being a stationary description, still knows the difference between past and future, must be imposed through a radiation condition. We use the procedure employed, e.g., by \citet[][\S 3.9]{lighthill78}, presuming that the external pressure has been turned on very slowly since $t=-\infty$
\be
  \hpext \to \pext \rme^{kV\epsilon t}, ~~ \epsilon = 0^+
\ee
where $\epsilon$ is defined to be dimensionless for convenience. 
s this transition amounts to the replacement rule $\kV \to \kV + \rmi kV \epsilon$. 
The tiny addition to the ship's ``frequency'', $\kV$, can be neglected except where it moves the pole slightly off the integration path, to complex values of $\bk$. 

The resulting expression for the surface elevation is now well defined and reads
\begin{align}\label{zetaGen}
    \zeta(\bxi) =& -\frac1{\rho}\lim_{\epsilon\to 0^+} \kint \frac{k\pext(\bk)\efac}{gk- (\bk\bcdot\bV)^2\coth kh -(\bk\bcdot \bV)(Sk_x/k)-\rmi\epsilon\Phi(\bk)},\\
    \Phi(\bk) =&  kV[2(\kV)\coth kh + Sk_x/k].
\end{align}
This result accords perfectly with that of \citet{havelock22} when $S=0$.
It is quite possible to use this expression directly for numerical purposes, keeping $\epsilon$ small but finite, as was done by \citet{moisy14}. The effect of $\epsilon$ is to attenuate the waves slightly away from the boat. In section \ref{sec:nearfar} we shall apply the Cauchy integral theorem as well as path of steepest descent techniques to obtain an expresseion for the far-field only. 

\subsubsection{Gaussian pressure source}
For definiteness, let us use the same Gaussian pressure source as used by \citet{ellingsen14a,ellingsen14b} and also by \citet{darmon14}:
\be\label{pext}
  \hpext(\bxi) = p_0 \rme^{-(\upi\xi/b)^2}; ~~ \pext(\bk) = \frac{b^2p_0}{\upi}\rme^{-(kb/2\upi)^2}.
\ee
As pointed out by \citet{he15}, such a model of a ship will not give a completely realistic scaling of the ``apparent wake angle'', the angle of maximum wave amplitude, for large Froude numbers, but this is not a point of focus in the present effort. It does, however, have the virtue of describing the size of the ``ship'' by a single parameter $b$, which in the presence of both shear and finite depth is seen as a great advantage for maintaining a manageable parameter space. 

It is well known, and recently shown explicitly for a model like ours with an anisotropic Gaussian external pressure ``ship'' model \citep{benzaquen14}, that the wave resistance depends on the shape of the ``ship''. Thus our wave resistance calculations in the following must be understood as a demonstration of the theory, while quantitatively accurate wave resistance calculation requires using a source $\hpext$ in Eq.~\eqref{Rgen} which approximates a particular hull shape. Such calculation is beyond the scope of the present effort. Note, however, that our formalism facilitates the use of more realistic models should quantitative results be needed in specific cases.

\subsection{Wave resistance}\label{sec:WRgen}

The theory for wave resistance on a travelling pressure disturbance was laid out long ago in a series of papers by \citet{havelock17,havelock19,havelock22}. This and other analytical models, as well as experimental results available at the time, were famously reviewed by \cite{wehausen73}. The effect of elongation of an ellipsoidal Gaussian pressure distribution was recently investigated by \citet{benzaquen14}. The effect of shear upon wave resistance has never been considered before to our knowledge.

Havelock's idea for calculation of wave resistance was to identify it as the horizontal component of the applied pressure force acting on the surface, in the direction of ship motion whereby the wave resistance may be found by an integral over the whole water surface:
\be
  R = 
  \int d^2\xi \,\hpext(\bxi) (V^{-1}\bV\bcdot\nabla_\xi)\zeta(\bxi).
\ee
where $\nabla_\xi = (\partial/\partial \xi_x, \partial/\partial \xi_y)$. Inserting Eq.~\eqref{zetaGen}, we recognise the complex conjugate of $\pext$ and find
\be\label{Rgen}
  R = -
  \frac{\rmi}{\rho V}
  \lim_{\epsilon\to 0^+} \kint \frac{k(\bk\bcdot\bV) |\pext(\bk)|^2}{gk- (\bk\bcdot\bV)^2\coth kh -(\bk\bcdot \bV)(Sk_x/k)-\rmi\epsilon\Phi(\bk)}.
\ee
The limiting expression in deep water is again given by letting $\coth kh\to 1$.

As shown by \citet{ellingsen14a} the presence of a shear current beneath the boat will result in an asymmetric wake when the angle $\beta$ is not $0$ or $\upi$. This, in turn, will give a lateral force on the wave source, normal to the direction of motion. In a slight misuse of terminology we term it the ``lateral wave resistance'' $R_\perp$, and it is found in analoguous fashion
\be
  R_\perp = \int d^2\xi \hpext(\bxi) [V^{-1}(\mathbf{e}_z\btimes\bV)\bcdot\nabla_\xi]\zeta(\bxi).
\ee
A positive value of $R_\perp$ means a lateral force directed towards the right (starboard) with respect to the ship's direction of motion. As for $R$ we find
\be\label{Rpgen}
  R_\perp = -\frac{\rmi}{\rho V} \lim_{\epsilon\to 0^+} \kint \frac{k[(\mathbf{e}_z\times\bV)\bcdot\bk] |\pext(\bk)|^2}{gk- (\bk\bcdot\bV)^2\coth kh -(\bk\bcdot \bV)(Sk_x/k)-\rmi\epsilon\Phi(\bk)}.
\ee

\subsection{Dispersion relation}\label{sec:disprel}

Much of the physics of any wave problem may be discerned from analysis of the dispersion relation. As discussed in \citet{ellingsen14a} (and numerous previous expositions without shear current present), the pole in the integrand of Eq.~(\ref{zetaGen}) --- that is, the zero of the denominator --- corresponds to values of $\bk$ which simultaneously satisfy the dispersion relation \emph{and} the condition of stationariness,
\be  \label{stationary}
  \kV = k c(\bk)
\ee
ensuring wave crests which look stationary as seen from the moving source. $c(\bk)$ is the phase velocity of a plane wave with wave vector $\bk$. 

Letting the denominator of Eq.~\eqref{zetaGen} (say) equal zero and solving with respect to $\kV$ immediately gives two solutions for the phase velocity,
\be\label{c}
  \kV = kc_\pm(\bk) = \pm\sqrt{gk\tanh kh +\left(\half S\cos\theta\tanh kh \right)^2}-\half S\cos\theta\tanh kh
\ee
where $\bk = (k\cos\theta,k\sin\theta)$. 
As discussed by \citet{ellingsen14b}, to each wave vector $\bk$ there are two associated phase velocities.
One of these phase velocities is positive, the other negative, and given the sign of $\kV$, the appropriate solution is chosen. 

For later reference let us denote the angle between $\bk$ and $\bV$ as $\gamma=\theta-\beta$, so that
\be
  \kV = kV\cos\gamma 
\ee
The consituents of the Fourier integral are plane waves with ``frequency'' $\kV$ which is negative when $|\gamma|>\upi/2$, and a wave of negative frequency and wave vector $\bk$ has phase velocity in the direction of $-\bk$. Thus the integral in Eq.~\eqref{zetaGen} (say) obtains two identical pole contributions, one from a wave of phase velocity $c_+$ whose wave vector has a forward component ($\kV>0$) and one of phase velocity $c_-$ whose wave vector has a rearward component ($\kV<0$). Both of these plane waves appear stationary as seen by the ship and give identical contributions. 

\begin{figure}
  \centerline{\includegraphics[width=\textwidth]{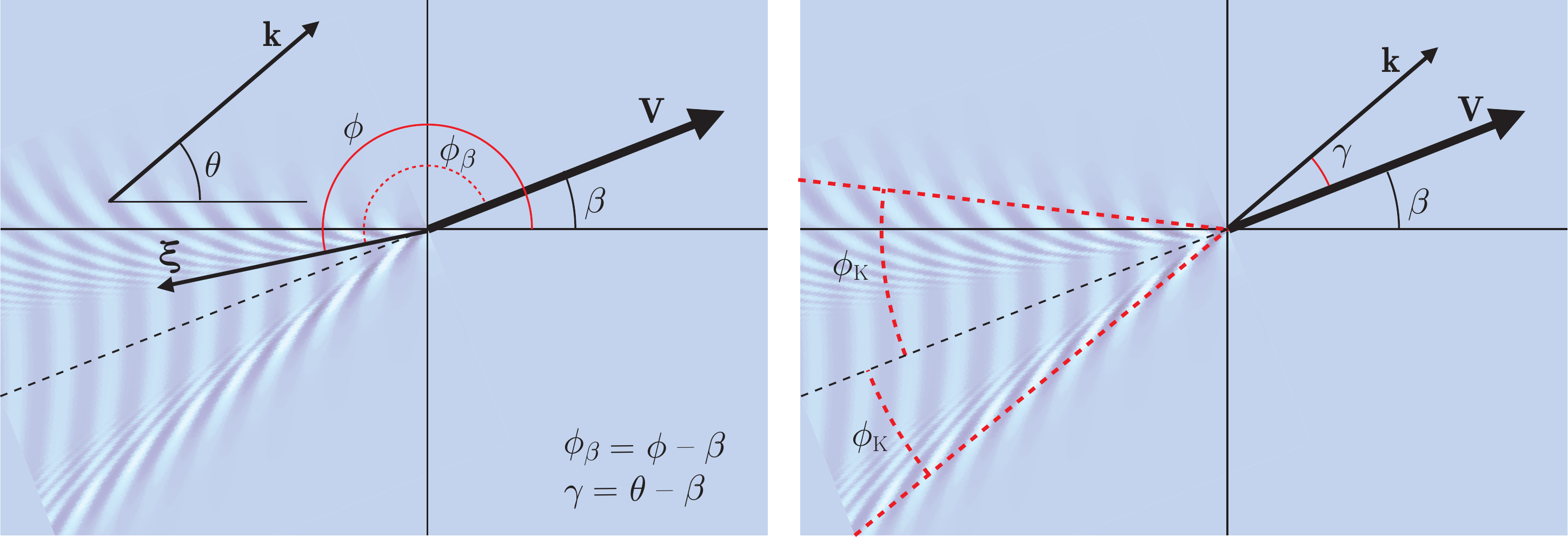}}
  \caption{Definitions of the different angles used in the analysis.}
  \label{fig:angles}
\end{figure}

While the co-ordinate system defined in Fig.~\ref{fig:geom} is easiest for the sake of the preceding derivations, for further analysis we shall want to use a polar system relative to the ship's motion, in the manner of \citet{ellingsen14a}. We define 
\be
  \phi_\beta = \phi - \beta. 
\ee
Using $\gamma$ and $\phi_\beta$ instead of $\theta$ and $\phi$ corresponds to rotating the coordinate system so that the boat moves along a new $x$ axis while the current, in general, does not. Definitions of the various angles used in the analysis are summarised in Fig.~\ref{fig:angles}.

\subsubsection{Length scales and limiting cases}

\newcommand{\lS}{l_\mathrm{S}}
Three length scales are involved in the model: $b$ (size of source), $h$ (depth) and a length associated with the shear, $\lS=g/S^2$. Known limiting cases are obtained if one of the length scales is much greater than the smaller of the other two. When $\lS\gg \min(h,b)$, the effect of shear becomes negligible,  waves are the same for all $\beta$, and the well known expressions of \cite{havelock08} are regained. Likewise, when $h\gg \min(\lS,b)$, the deep water case considered by \cite{ellingsen14a} is found, and the simplest case $b\ll h,\lS$ is detailed in \cite{darmon14}, being the deep water case with no shear. 

The most pertinent physical insights are obtained when these length scales are compared to typical wavelengths in different parts of the wake, as obtained from stationary phase arguments. Once typical values of $k$ and $\theta$ in some part of the wake (transverse or diverging waves), limiting cases can be analysed using the dispersion relation \eqref{c}, which reveals that the effect of shear is weak provided $\delta\equiv \tanh kh/(k\lS)\ll 1 $. In this case
\be 
  c_\pm(\bk) = c_0(k)\left(\pm 1 - \frac{1}{2}\sqrt{\delta}\cos\theta \pm \frac18 \delta\cos^2\theta  +...\right)
\ee
with $c_0(k)=\sqrt{(g/k)\tanh kh}$. We see that for the effect of shear upon a wave $\bk$ to be small, it is sufficient that $|\sqrt{\delta}\cos\theta|\ll 1$. There are thus two cases in which shear is rendered unimportant even though $k\lS$ is not large: if $\theta$ is close to $\pm\pi/2$ (propagation normal to the shear current), or if $kh\ll 1$ (shallow water waves). The relative unimportance of shear for shallow water waves was shown for ring waves by \cite{ellingsen14b}.

\subsection{Near-field and far-field contributions}\label{sec:nearfar}

We wish now to extract the far-field contribution to the ship waves. The analysis is carried out in a detailed and careful manner, thereby exposing weaknesses in a more cavalier method used in the recent literature, including \cite{darmon14} and \cite{ellingsen14a}. In order not to clutter the reading overly with mathematics, some calculations are found in appendices.

We will re-write the expression for the surface elevation \eqref{zetaGen} with a Gaussian disturbance in a different form which is suitable for further analysis. First, let us write the integral over $\bk$ in polar form with dimensionless quantities,
\bs \label{zetanondim}
\begin{align}
  \zeta(\bX) =& \frac{bp_0}{4\upi^3\rho V^2}\pipiint \frac{\rmd\gamma}{\cos^2\gamma}\lim_{\epsilon\to 0^+}I(\gamma) \\
  I(\gamma) =& \Kint K\frac{\rme^{E(K,\gamma)}\tanh KH }{\Gamma(K,\gamma) + \rmi\epsilon \Psi(K,\gamma)} \equiv \Kint f(K,\gamma)
\end{align}
\es
where we have defined
\be
  \bK = b\bk, ~~ \bX = \bxi/b, ~~ H = h/b.
\ee
and the functions (for frequent reference below)
\bs
\begin{align}
  E(K,\gamma) =& -K^2/4\upi^2+\rmi KX\cos(\gamma-\phi_\beta) \label{E},\\
  \Gamma(K,\gamma) =& K - \frac{f_s(\gamma)}{\Fr^2 \cos^2\gamma}\tanh KH, \\
  \Psi(K,\gamma)=& \frac{2K}{\cos\gamma} + \frac{\Frs}{\Fr^2}\frac{\cos(\gamma+\beta)}{\cos^2\gamma}\tanh KH, \\
  f_s(\gamma) =& 1-\Frs \cos\gamma\cos(\gamma+ \beta).\label{fs}
\end{align}
\es

Our  system is described by four nondimensional parameters: $\Fr, \Frs,\Frh$ and $\beta$, where the three Froude numbers are
\be\label{froude}
  \Fr = \frac{V}{\sqrt{gb}};~~ 
  \Frs = \frac{VS}{g}; ~~ 
  \Frh = \frac{V}{\sqrt{gh}}.
\ee
Here $\Fr$ is based on the size of the source (the ship), $\Frs$ is based on a ``shear depth'' $g/S^2$ which is half the depth at which the dynamic and hydrostatic pressure of the shear flow are equal, and $\Frh$ is based on the finite water depth.

\begin{figure}
  \centerline{\includegraphics[width=\textwidth]{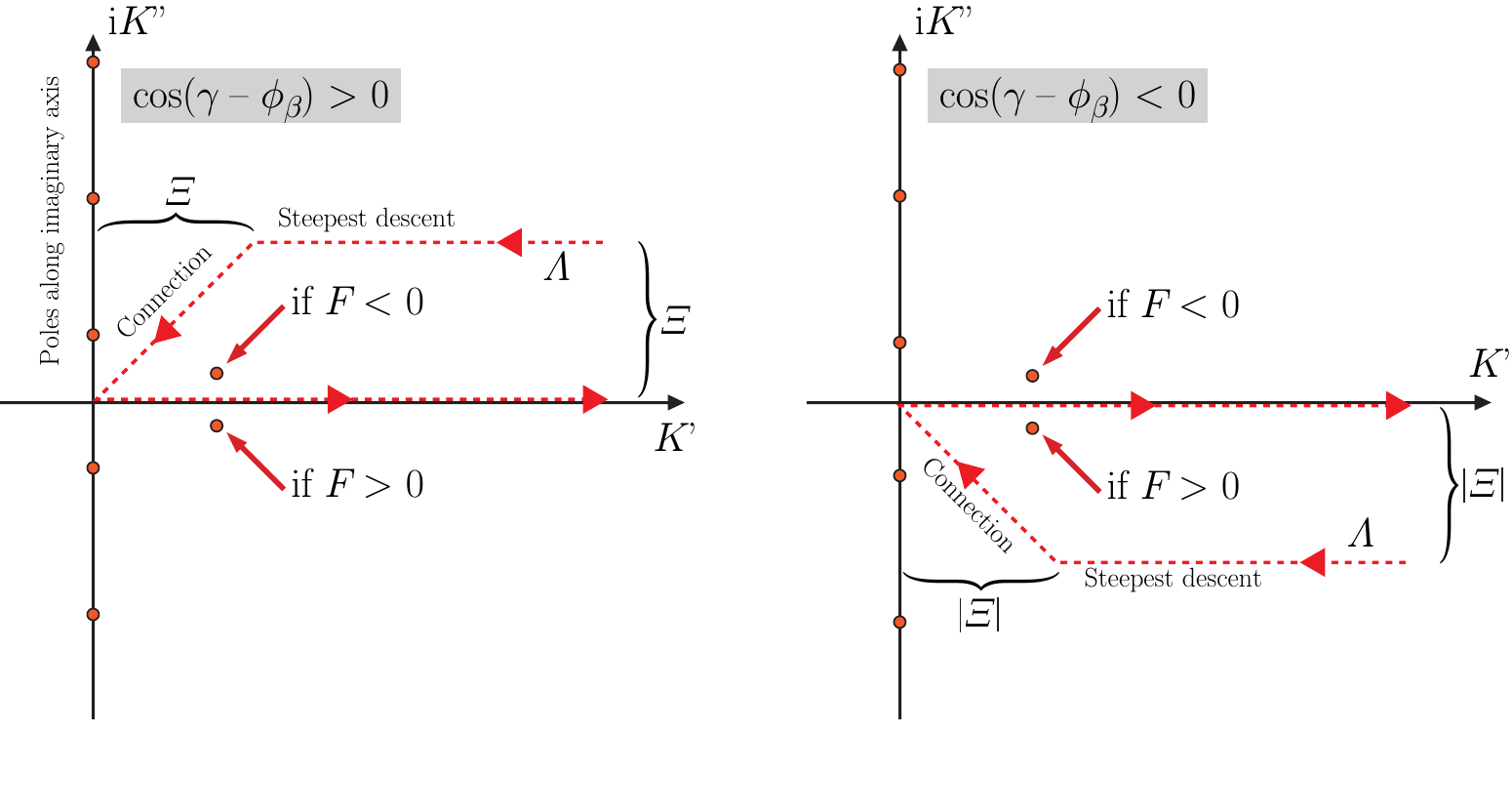}}
  \caption{Contours of integration: the original integration path along the positive real $K$ axis is closed with a steepest descent path and a connection path at an angle with both axes. The position of the pole is indicated, its position relative to the real $K$ axis depending on the sign of the function $F(K_0,\gamma)$.}
  \label{fig:contours}
\end{figure}

We will consider the integral $I(\gamma)$ by forming a closed contour in the complex $K$ plane. The contour is formed of the positive real $K$ axis (the original integration path) and closed either in the upper or lower plane, depending on the exponent function $E(K,\gamma)$ so that the resulting path gives a finite and well defined integral. To wit we shall choose the path of steepest descent \citep[c.f., e.g.][\S 6.6]{bender91}, 
\be
  K_\text{s.d.}(K,\gamma) = K +  2\upi^2 \rmi X\cos(\gamma-\phi_\beta).
\ee
This path is parallel to the real $K$ axis and lies either above or below the latter depending on the sign of $\cos(\gamma-\phi_\beta)$. We connect it to the original path of integration while noticing that $\Gamma(K,\gamma)$ has a series of zeros along the imaginary axis which we wish to avoid, and therefore choose the connection path (arbitrarily) at $45^\circ$ to the real axis. The path is closed at infinity where the integrand is exponentially zero. The closed path of integration, $\Lambda$, is shown in figure \ref{fig:contours}.

Now we define $K_0(\gamma)$ as the (real) value of $K$ for which $\Gamma(K,\gamma)$ has a zero, i.e., it is implicitly defined by
\be\label{K0def}
  K_0 -  \frac{f_s(\gamma)}{\Fr^2 \cos^2\gamma}\tanh K_0H =0.
\ee
Only in the deep water limit $H\to\infty$ is the expression for $K_0$ explicit. For a given set of paramerers, it is not certain that a positive solution of Eq.~\eqref{K0def} exists. Note that the trivial solution $K_0=0$ is not a pole of the integrand since it is cancelled by a factor $K$ in the numerator. The existence or non-existence of a positive solution $K_0(\gamma)$ is related to the question of a critical velocity, and is discussed in Section \ref{sec:crit}. There, an approximate, explicit solution for $K_0(\gamma)$ is also given.

Importantly, when $K_0$ is inserted for  $K$, the exponent function $E(K_0,\gamma)$ does not depend on the source size $b$, i.e., on the Froude number $\Fr$. This is important because the pole at $K=K_0$ gives far-field waves (to be shown below), and the exponent function $E(K_0,\gamma)$ is what determines the width of the Kelvin wake. Hence, just as was for deep water \citep{ellingsen14a} (and is well known to be the case without shear), the source Froude number has no influence on the Kelvin angle, although it strongly affects the \emph{apparent} wake angle at which the waves have the largest amplitude \citep[see, e.g.][]{darmon14,noblesse14}.

\begin{figure}
  \begin{center}
    \includegraphics[width=\textwidth]{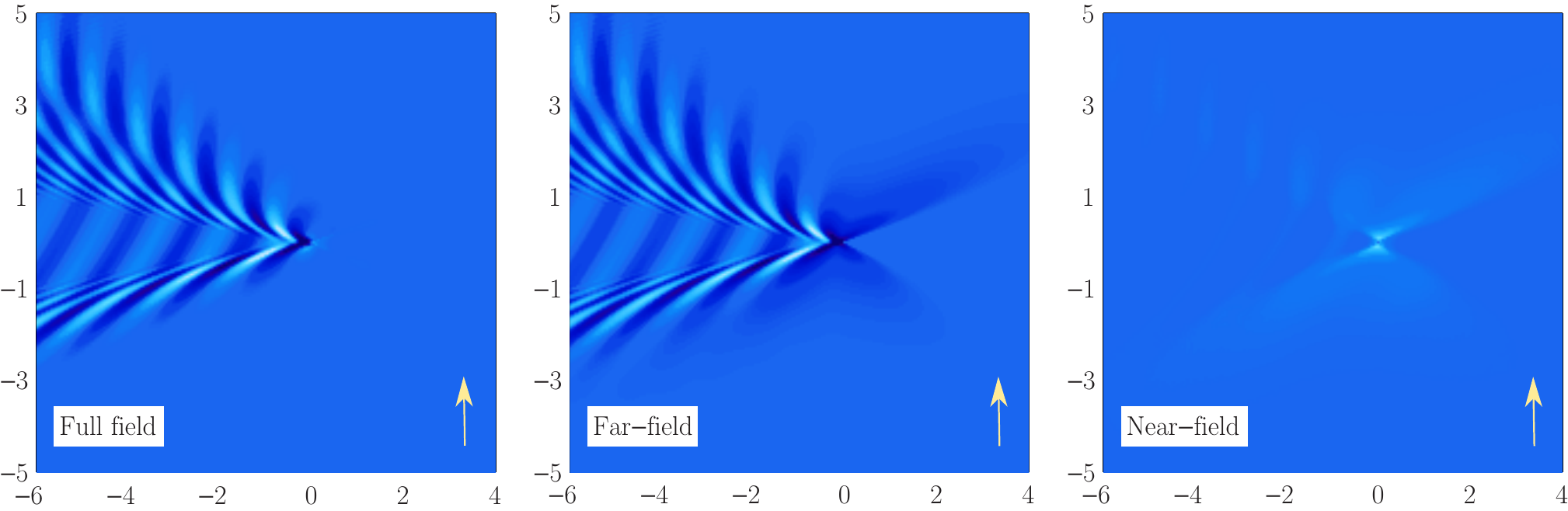}
  \end{center}
  \caption{Illustration of wake waves calculated using the expressions for (left) the full expression in Eq.~\eqref{zetaGen}, (centre) the far-field expression in Eq.~\eqref{zetafar}, and (right) only the near-field. In all panels $\Fr=\Frs=0.8$, $\Frh=0$ and $\beta=\upi/2$. Arrows indicate direction of shear flow as defined in Fig.~\ref{fig:geom}.}
  \label{fig:nearfar}
\end{figure}

We now proceed to solve the contour integral $I(\gamma)$. We show in Appendix \ref{app:nearfield} that the integrals $\Isd$ and $\Ic$ make up the near field of the wake, i.e., a surface deformation following the source which falls off quickly as $X$ increases. The all-important far field of the wave pattern is given by the contribution from the pole, provided it is found inside the contour. This should be no surprise, since exactly this decomposition has been reported numerous times in the literature \citep[e.g.][]{lighthill78} for ship waves as well as other wave systems. 

Evaluating the contribution from the pole (details may be found in Appendix \ref{app:polecontrib}) we obtain the following expression for the far-field of the wave pattern
\begin{align}\label{zetafar}
  \zff =& -\frac{\rmi p_0}{2\upi^2\rho g}\pipiint \rmd\gamma\Theta(K_0)\Theta[-\cos(\gamma-\phi_\beta)\cos\gamma]\sg[\cos\gamma]\notag \\
  &\times\frac{K_0\rme^{E(K_0)}\tanh K_0H}{\Fr^2\cos^2\gamma-Hf_s(\gamma)\sech^2K_0H}.
\end{align}
Here $\Theta$ is the Heaviside step function.

Crucially, once the far-field is identified as the contribution to $I(\gamma)$ from the pole, it implies that a wave of propagation direction $\gamma$ contributes to the far field in real-space direction $\phi$ if and only if the pole where $\Gamma+\rmi\epsilon\Psi=0$ lies inside the contour $\Lambda$. Much of the below analysis rests upon this insight, and we shall see that the two different ways in which the pole can fall outside the contour, corresponding to the two Heaviside functions in Eq.~\eqref{zetafar}, each have their different physical interpretations. 

The separation into near-field and far-field are shown in Fig.~\ref{fig:nearfar}. One may note that the far-field as calculated with the Cauchy integral theorem gives a butterfly-like surface deformation near the source, which is particularly visible with strong side-on shear such as shown in Fig.~\ref{fig:nearfar}. The artifact is not a worry since the far-field expression is only accurate far from the source. 

The requirement that the pole most lie inside the closed contour in order to contribute led to the Heaviside factor $\Theta[-\cos(\gamma-\phi_\beta)\cos\gamma]$ in Eq.~\eqref{zetafar}. A careful inspection of the angles involved reveals that this factor restricts the contribution to the far-field to including only the partial waves whose direction of propagation has a positive component towards the ship. In other words, the waves in the far-field always \emph{follow} the ship and are not allowed to run ahead of it. This accords well with what one must expect, but we note that arguments based on the group velocity, say, would be complicated since in the presence of shear the phase and group velocities do not in general have the same direction. 

The radiation condition used here automatically allows for the perhaps surprising result that with side-on shear near the critical velocity, waves can in fact be seen in front of the moving ship on one side (see section \ref{sec:kelvin} and \citep{ellingsen14a}). This is not in violation of the radiation condition because these waves have been sufficiently refracted by the shear to still be able to follow the moving source according to the above definition.

Note finally that, beyond being a generalization, the far field expression in Eq.~\eqref{zetafar} in fact differs slightly from those reported in \citet{ellingsen14a, darmon14,benzaquen14}. In these references the Sokhotsky--Plemelj formula was used to extract the far-field contribution from an expression similar, but not identical, to Eq.~\eqref{zetaGen}. We explain in appendix \ref{app:SP} why the use of this theorem is trecherous; it yields a far-field expression which but for a factor $2$ is identical to ours in the limit $X\to \infty$ \emph{behind} the ship, but which contains a spurious wake also \emph{in front} of the ship which is equal but of opposite sign. Clearly such a far-field does not satisfy the radiation condition at positions within the spurious Kelvin wedge in front of the ship, the reason for which appendix \ref{app:SP} elucidates. When Sokhotsky--Plemelj is used, the spurious wake must be manually removed from the far-field expression, for instance by simply not plotting it. This is easy and clear cut in deep waters with no shear current, when the wave train is famously contained within a Kelvin wedge with half-angle $19^\circ 28'$, but not so straightforward when the wake angle grows large or even extends beyond $90^\circ$. Moreover, \citet{ellingsen14a} and \citet{darmon14} perform integration over $\gamma$ only from $-\upi/2$ to $\upi/2$, resulting in a further factor $2$ difference and afar-field expression which is an overall factor $4$ smaller than our Eq.~\eqref{zetafar}. This does not affect any of the main conclusions in these references.

\section{Critical velocity}\label{sec:crit}

In this section we discuss the phenomenon of \emph{critical velocity}, and derive the criterion for criticality, Eq.~\eqref{supercrit1}, when both shear and finite depth are present. 

The phenomenon of critical velocity is known previously both for ship waves in shallow water \citep{havelock08} and in shear current \citep{ellingsen14a}. In physical terms, when the ship's speed exceeds a certain critical velocity, transverse-propagating waves (i.e., the part of the ship waves whose direction of propagation $\bk$ is close to parallel with $\bV$, found directly behind the ship in a wedge including $\phi_\beta=\upi$) are unable to keep up with the source and cannot contribute to a stationary wake as seen by the moving source. We will show that the criterion that $K_0(\gamma)$ must exist for waves contributing to the far-field is exactly the criterion which ensures that transverse plane wave components are excluded at supercritical velocities.

Below we will derive the following explicit expression for the critical Froude number and velocity,
\be\label{FrC}
  \Fr_\text{crit}=\frac{\Vc}{\sqrt{gb}} = \frac{\sqrt{\Frsb^2+1/H}-\Frsb\cos\beta}{1/H+\Frsb^2\sin^2\beta}
\ee
where the intrinsic Froude number (based on the velocity $Sb$ and the source size $b$) is 
\be\label{Frsb}
  \Frsb = \frac{Sb}{2\sqrt{gb}} = \frac{S}{2}\sqrt{\frac{b}{g}}.
\ee

While it is true that the amplitude of the transverse waves in the wake decreases as the ship's velocity (hence $\Fr$) increases, and they gradually vanish from sight \citep{darmon14}, the vanishing of transverse waves is not the only, or even the most striking phenomenon associated with the critical velocity. As $V$ approaches the critical, the total Kelvin wake angle reaches $\upi$ in a sharp peak at this velocity, as will be duly discussed in the following. The phenomenon can not be observed in deep, still waters since it is caused by the fact that phase velocity is limited: In shallow waters, the phase velocity cannot exceed $\sqrt{gh}$, and in a shear current $c(\bk)$ is limited above by $g/S\cos\theta$ when $\theta>0$ (propagation against the shear) as may be deduced from Eq.~\eqref{c}.

As argued, the value of the wave number which satisfies both the dispersion relation and the condition that the wave front appears stationary as seen from the boat is $K_0(\gamma)$, solving Eq.~\eqref{K0def}. Only such a plane wave can contribute to the ship wave pattern in the far-field. It is useful to write Eq.~\eqref{K0def} in the form
\be\label{K0B}
  f(\varkappa)\equiv \varkappa - B(\gamma)\tanh \varkappa = 0
\ee
with $\varkappa = K_0H$ and $B(\gamma)=f_s(\gamma)/(\Frh^2\cos^2\gamma)$. For any real $B(\gamma)$ the function $f(\varkappa)$ is smooth, positive as $\varkappa\to\infty$ and is zero at $\varkappa=0$, hence it follows that $f(\varkappa)$ has a zero at some positive $\gamma$ if and only if $f'(0)<0$, that is, if $B>1$. Thus the criterion for a positive solution $K_0(\gamma)>0$ to exist for some value of $\gamma$ is that $B(\gamma)>1$,
which can be written
\be\label{critGamma}
  1-\Frs\cos\gamma\cos(\gamma+\beta)-\Frh^2\cos^2\gamma >0.
\ee

An approximate solution to \eqref{K0B} for $\varkappa$ as a function of $\gamma$ expressed as a functional of $B(\gamma)$ may be found by matching the asymptotic behaviours at $B\to \infty$ and $B\to 1$, 
\be\label{kapp}
  \varkappa_\mathrm{app} = B(\gamma) - 1 + \sqrt{1-B(\gamma)^{-3}}.
\ee
The approximation remains better than approximately $8$\% accurate for all $B>1$, and is $1\%$ accurate or better for $B\gtrsim 2$.

\subsection{Condition of criticality}

\begin{figure}
  \centerline{\includegraphics[width=3in]{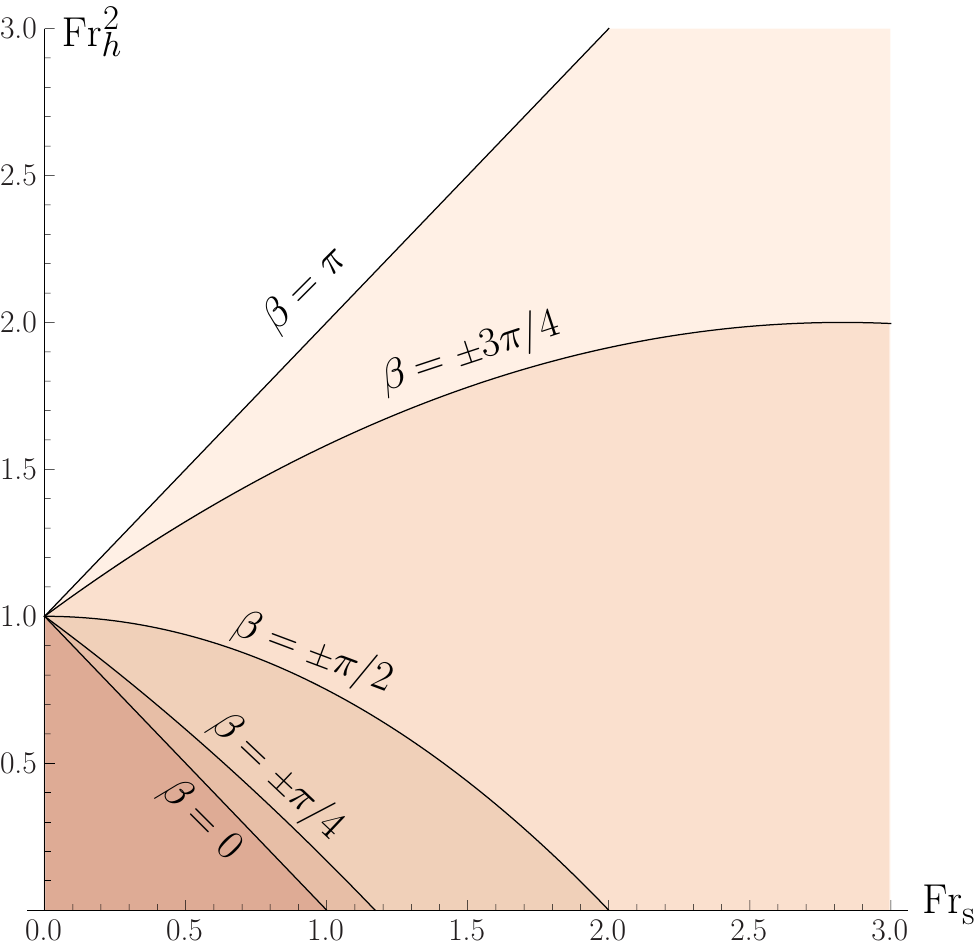}}
  \caption{Critical lines as function of $\Frs$ and $\Frh$ for different values of $\beta$. The shaded regions below the critical lines are sub-critical.}
  \label{fig:critcurve}
\end{figure}

We will now derive and discuss the critical velocity in the present case in which both finite depth (parameterised though $\Frh$) and shear current (parameterised through $\Frs$) are present. Either of the two on its own will give rise to a finite critical velocity, and combining the two naturally yields a critical velocity which depends on both $\Frh$ and $\Frs$, as well as the angle $\beta$ between ship motion and shear current. 

We shall describe a situation given by parameter triplet $\Frh,\Frs,\beta$ as \emph{supercritical} if a finite sector of $\gamma$ values exists wherein the criterion \eqref{critGamma} is false. It is slightly easier mathematically to work with the re-arranged condition, equivalent to \eqref{critGamma},
\[ 
    \frac{1-\Frs\cos\gamma\cos(\gamma+\beta)}{\cos^2\gamma} >\Frh^2
\]
(we presume $\cos\gamma\neq 0$). The situation is supercritical if 
\[
  \min_\gamma \Bigl\{\frac{1-\Frs\cos\gamma\cos(\gamma+\beta)}{\cos^2\gamma}\Bigr\}<\Frh^2
\]
where the notation means the minimum of the left-hand side with respect to $\gamma$ is taken. The minimum is found at $\tan\gamma=-\half\Frs\sin\beta$, which we reinsert and conclude that the situation is supercritical if 
\be\label{supercrit1}
  \Frs(\cos\beta+\quarter\Frs\sin^2\beta) + \Frh^2 > 1.
\ee
The sub- and supercritical regions of the $\Frs$-$\Frh$ parameter plane are shown in Fig.~\ref{fig:critcurve} for some values of $\beta$.

\begin{figure}
  \includegraphics[width=\textwidth]{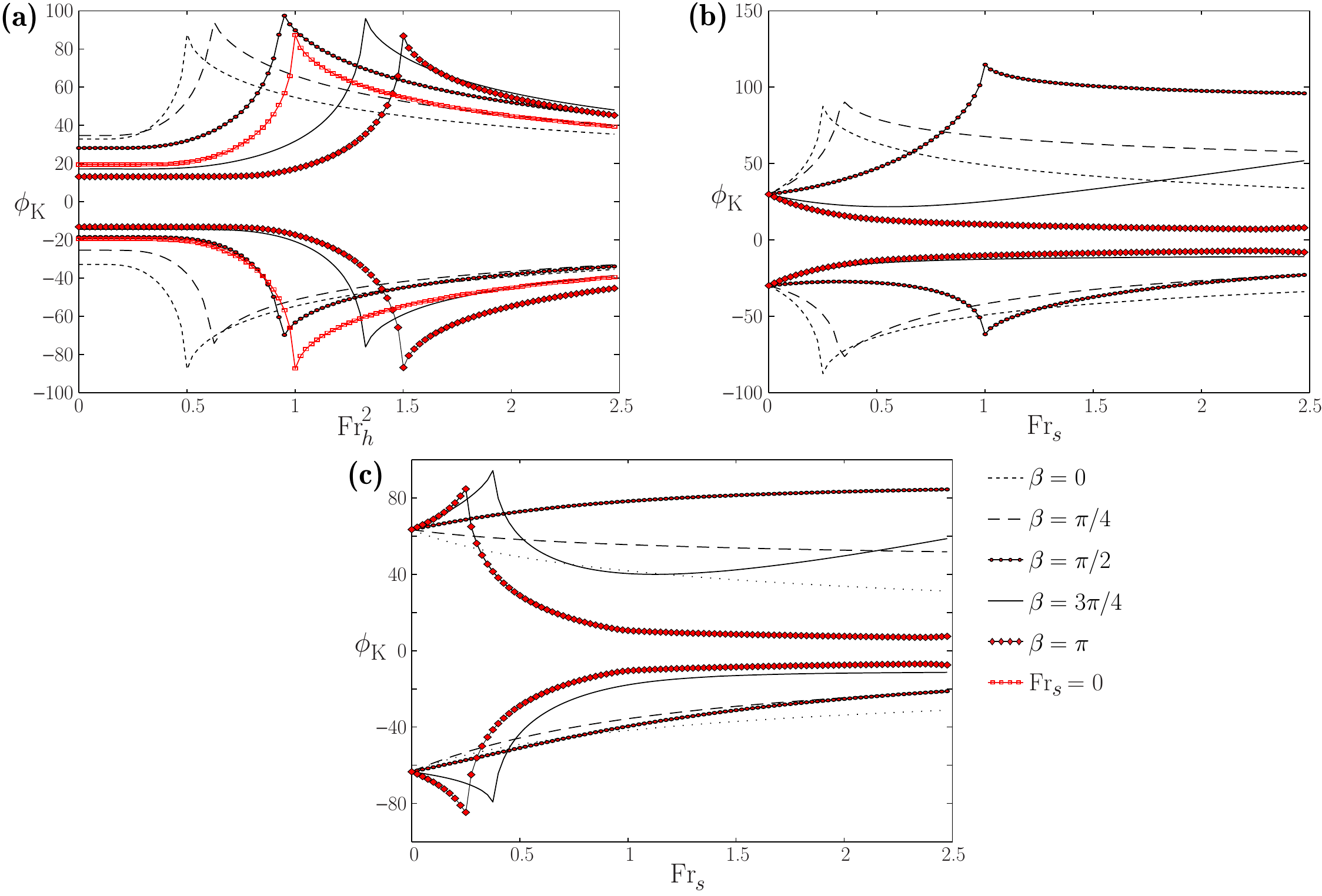}
  \caption{Kelvin angles for constant $\Frs=0.5$ (a), constant $\Frh^2=0.75$ (b) and constant $\Frh^2=1.25$ (c). The case $\Frs=0$ is also shown in the panel (a) for comparison. See also Fig.~\ref{fig:kelvinpanels}.}
  \label{fig:kelvinplots}
\end{figure}

\begin{figure}
  \includegraphics[width=\textwidth]{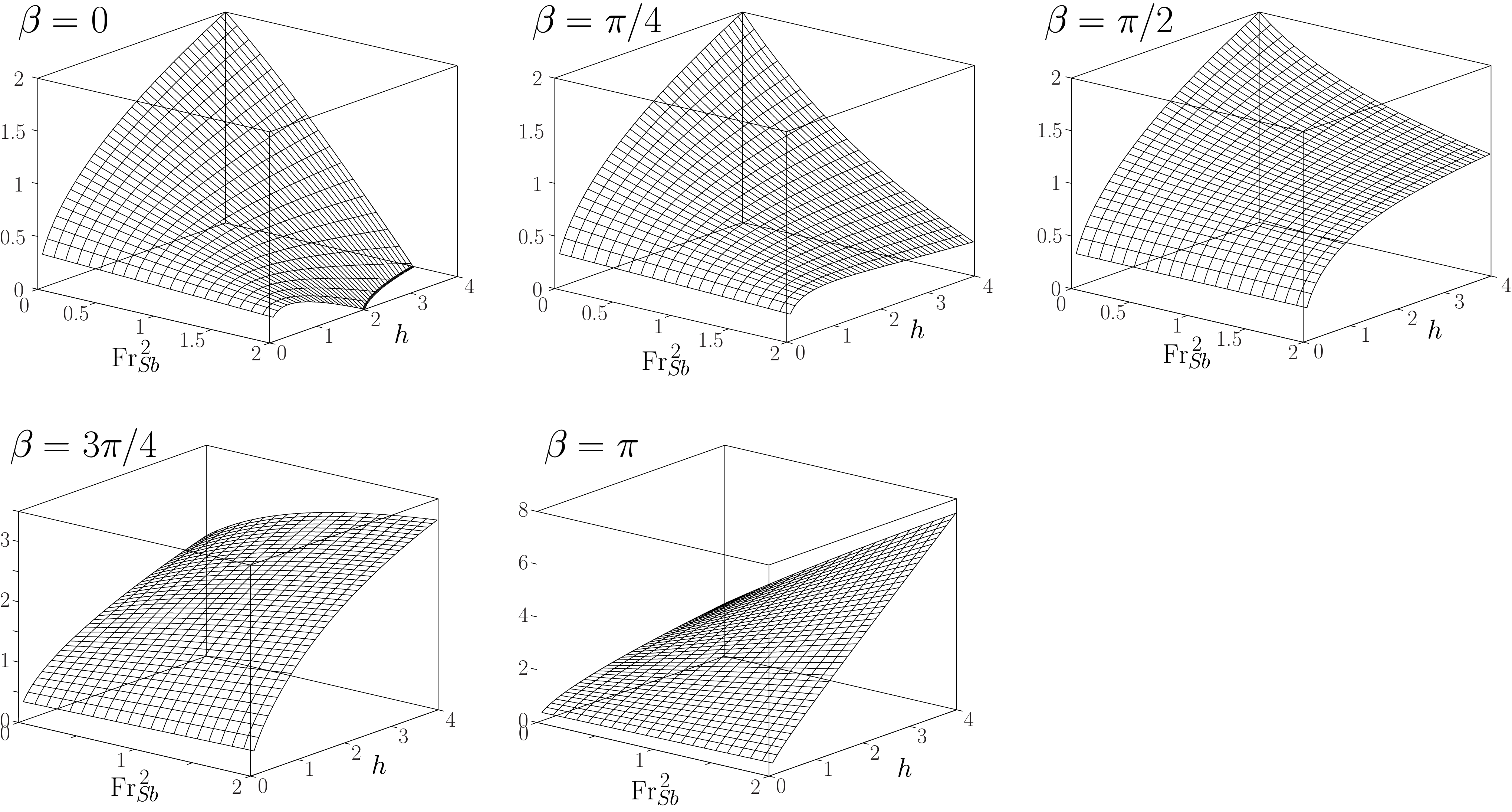}
  \caption{$\Fr_\text{crit}$ as a function of $h$ and $\Fr_{Sb}$ for different values of $\beta$.}
  \label{fig:Frcrit}
\end{figure}

Inserting the definitions of $\Frs$ and $\Frh$, Eq.~\eqref{supercrit1} can be solved with respect to $V$ to obtain the critical velocity given in Eq.~\eqref{FrC}. 
The known limits when $S=0$ (no shear current) when $\Vc=\sqrt{gh}$ \citep{havelock08}, and as $h\to\infty$ when $\Vc=S/g\cos^2(\beta/2)$ \citep{ellingsen14a}, are regained. The latter limit may be seen easily if one notes that Eq.~\eqref{supercrit1} can instead be written
\be\label{supercrit2}
  \frac{\Frh^2}{1+\Frs\sin^2(\beta/2)} + \Frs \cos^2(\beta/2) > 1.
\ee
We plot $\Fr_\text{crit}$ in Fig.~\ref{fig:Frcrit}.

\section{Wave patterns and Kelvin angles under different conditions}\label{sec:kelvin}

In this section we present numerical evaluation of wave patterns and Kelvin angles (total angular extent of the wake) in different parts of the parameter space spanned by parameters $\Fr,\Frs, \Frh$ and $\beta$. 

Combining what has been found so far, the far-field of the surface elevation can be written
\begin{align}\label{zetafar2}
  \zff =& \frac{p_0}{2\upi^2\rho g}\pipiint \rmd\gamma\Theta[1-\Frs\cos\gamma\cos(\gamma+\beta)-\Frh^2\cos^2\gamma]\sg[\cos\gamma]\notag \\
  &\times\Theta[-\cos(\gamma-\phi_\beta)\cos\gamma]\frac{K_0\rme^{-(K_0/2\upi)^2}\sin[K_0X\cos(\gamma-\phi_\beta)]\tanh K_0H}{\Fr^2\cos^2\gamma-Hf_s(\gamma)\sech^2K_0H}
\end{align}
where we have now taken the real part. Significant contributions to the far--field are only obtained for values of $\phi_\beta$ where the argument of the sine,
\be
  f_1(\gamma) = K_0(\gamma)X\cos(\gamma-\phi_\beta)
\ee
has a stationary point for a value of $\gamma$ in the integration range, that is, where $\partial_\gamma f_1(\gamma)=0$. The Kelvin angle as defined in Refs.~\citet{darmon14} and \citet{ellingsen14a} is the largest value of $|\upi-\phi_b|$ for which a stationary point exists; in the presence of a shear current the Kelvin angle is in general different on either side of the wake.

Unlike for the cases considered in these references we do not now have an explicit expression for $K_0(\gamma)$ or its derivative, so the Kelvin angle must be found numerically by first calculating the value of $\phi_{\beta,\text{stat}}(\gamma)$ at which a stationary point exists for a given value of $\gamma$ and then finding the extrema of $\upi-\phi_{\beta,\text{stat}}(\gamma)$ in the range of $\gamma$ \citep[see also][for details and illustration]{ellingsen14a}. 

\afterpage{%
\begin{landscape}
  \begin{figure}
    \includegraphics[height=.9\textwidth]{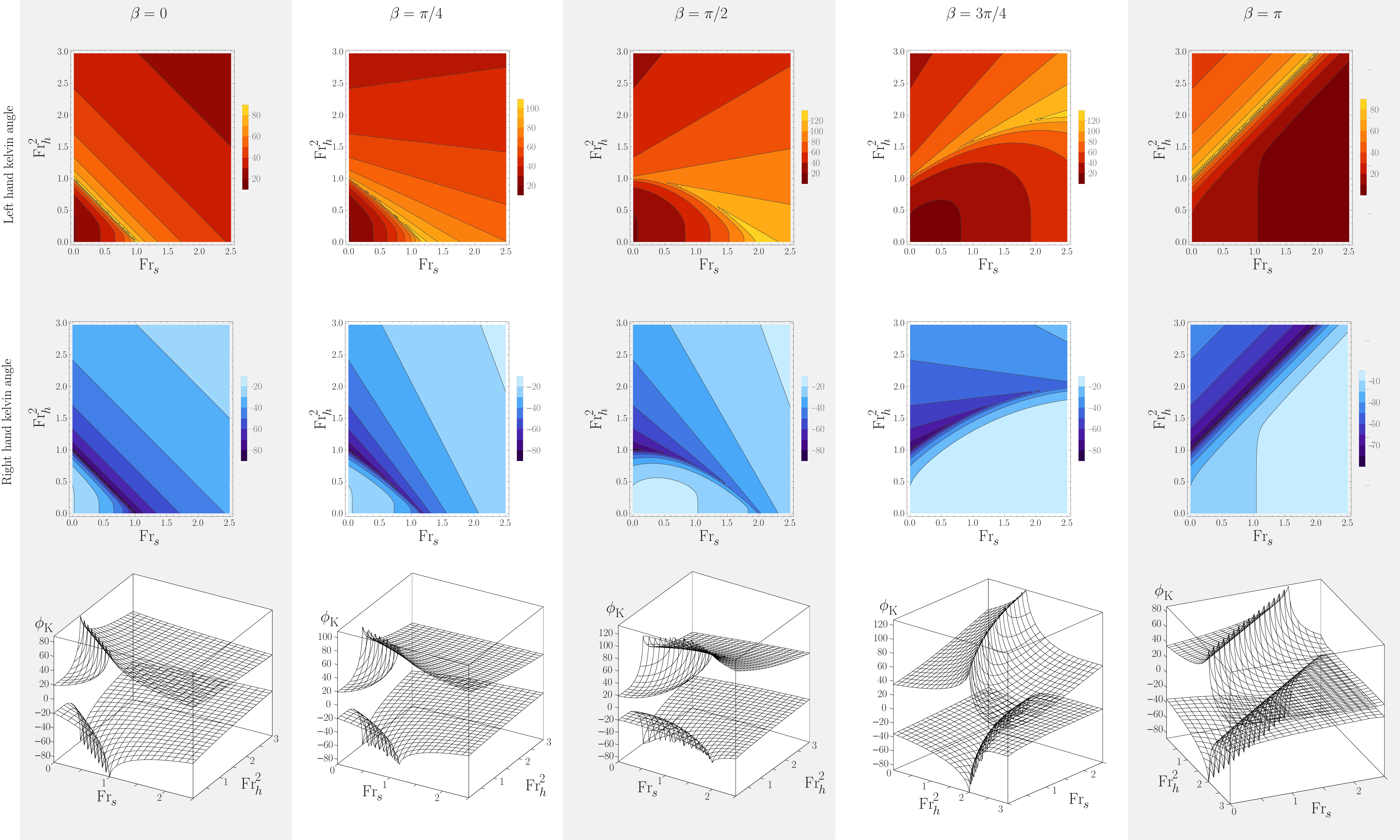}
    \caption{Kelvin angles for different values of $\Frs$ and $\Frh^2$. See also Fig.~\ref{fig:kelvinplots}.}
    \label{fig:kelvinpanels}
  \end{figure}
\end{landscape}
}

The Kelvin angles for different values of parameters $\Frs$ and $\Frh^2$ are shown in Fig.~\ref{fig:kelvinplots} and Fig.~\ref{fig:kelvinpanels}. As is clear to see, the Kelvin angles on both sides of the wake show sharp maxima on the critical curves in the $\Frs$-$\Frh^2$ plane which we plotted in Fig.~\ref{fig:critcurve}. Exactly as for deep waters, the Kelvin angle is not influenced by $\Fr$. A very pronounced effect is, however, that as $\Fr$ exceeds $1$ the wake \emph{appears} to grow narrower with increasing $\Fr$ because the largest wave amplitudes are found at wake angles smaller than the Kelvin angle with an \emph{apparent} angle decreasing as $\Fr^{-1}$ for the Gaussian source as discussed in the Introduction. When $\Fr$ is large, neither the presence of a shear current \citep{ellingsen14a} nor finite depth \citep{zhu15} have more than a modest effect on the apparent wake angle, however, and considering the already large parameter space of our model we shall not focus on the effects of varying $\Fr$ in the present effort.

\begin{figure}
  \centerline{\includegraphics[width=\textwidth]{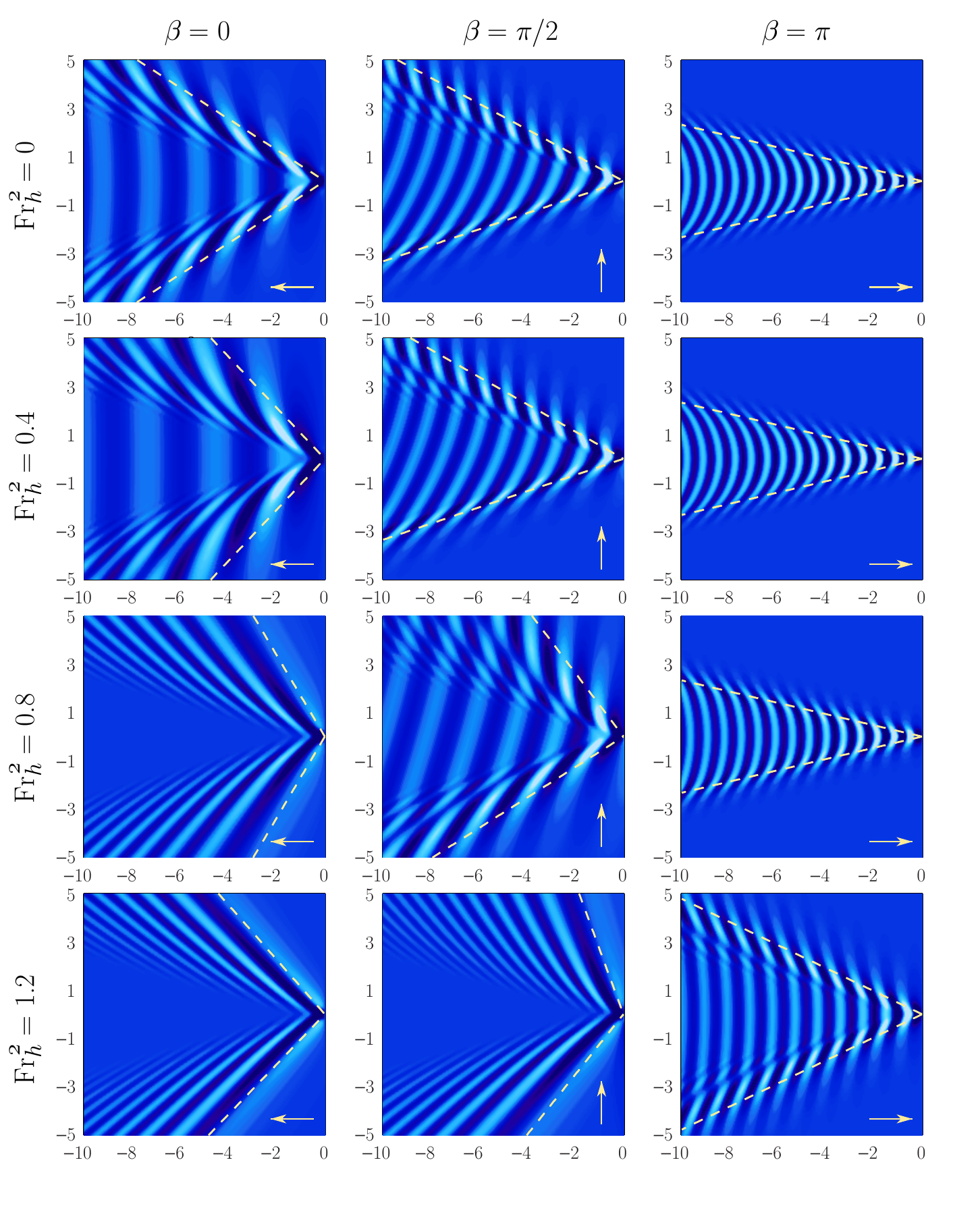}}
  \caption{Wave patterns evaluated from the far-field expression Eq.~\eqref{zetafar2}, for different water depths and angles $\beta$ of ship motion relative to the shear current. Length scales are in units of $2\upi b\Fr^2$. In all graphs, $\Fr=0.5$ and $\Frs=0.5$. A comparison with Fig.~\ref{fig:critcurve} shows that the two bottom panels in the left column and the bottom middle panel show supercritical situations. Arrows indicate direction of shear flow in the system where the surface is at rest (see Fig.~\ref{fig:geom}).}
  \label{fig:panels}
\end{figure}

\begin{figure}
  \centerline{\includegraphics[width=\textwidth]{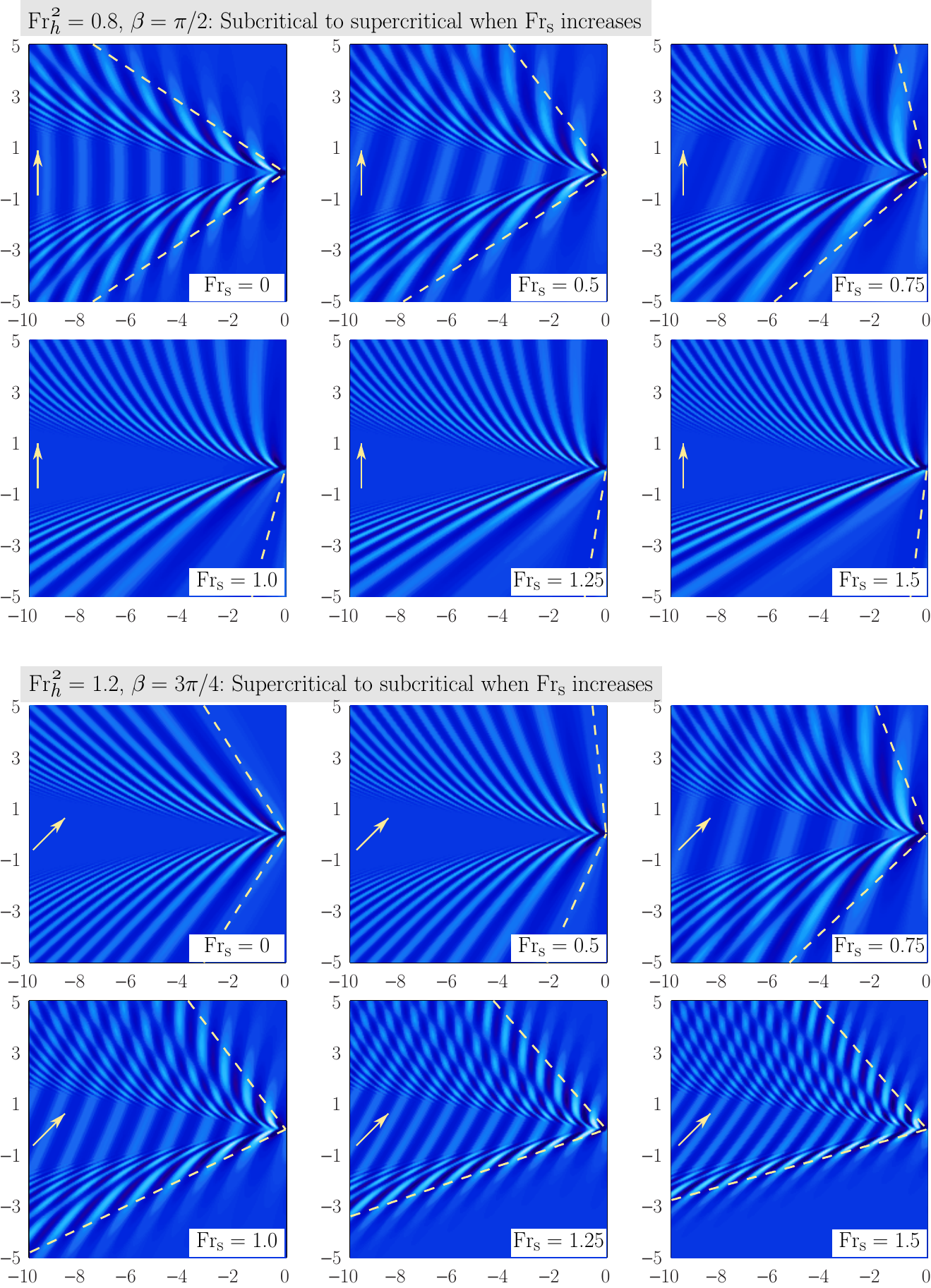}}
  \caption{Increasing the shear $S$ (and therefore $\Frs$) can cause transition from subcritical to supercritical situation (top 6 panels) or from supercritical to subcritical (bottom 6 panels) depending on the value of $\Frh$ and $\beta$. In deep waters ($\Frh= 0$) only the former transition is possible. In all graphs $\Fr=0.8$. Arrows indicate direction of shear flow in the system where the surface is at rest (see Fig.~\ref{fig:geom}).}
  \label{fig:transition}
\end{figure}

The effect of varying depth on the ship waves is illustrated in Fig.~\ref{fig:panels} where the wave field is shown for different values of $\Frh^2$ when $\Fr$ and $\Frs$ are held constant at moderate values. The values of $H$ for each row (top to bottom) are $\infty,1.6,0.8$ and $0.53$. As the depth decreases the waves for $\beta=0$ (upstream) and $\beta=\upi/2$ (side-on) both become supercritical, with transverse waves disappearing. The presence of the sea bed is felt most strongly for upstream ship motion ($\beta=0$), whereas for downstream ship motion ($\beta=\upi$) the effect of finite depth only becomes noticeable for the largest $\Frh$, when the wake goes from being  made up purely of transverse waves to also showing diverging contributions as well as a wider Kelvin angle. The waves following the downstream-going source are helped along by the current, shortening the wavelength required for transverse waves' velocity to equal that of the source. A plane wave is affected by the sea floor only if depth is less than roughly half a wavelength, and due to the shortened wavelength, at $\beta=\pi$ the sea bed only begins to be felt at values of $\Frh^2$ which are supercritical in the absence of shear.

That a shear current can cause transition from subcritical to supercritical waves when the shear $S$ is increased was already shown by \citet{ellingsen14a}. However, as Fig.~\ref{fig:critcurve} shows, at finite depth it is also possible to effect the opposite transition by increasing the shear. We illustrate this in Fig.~\ref{fig:transition}. In the top 6 panels of the figure, the normal transition is seen from sub- to supercritical when $S$ is increased with other quantities constant (in non-dimensional terms, $\Frs$ is increased with constant $\Frh$ and $\Fr$). Here the direction of motion is $\beta=\upi/2$ and $H \approx 0.31$, and $\Fr=0.5$. Now changing the direction of travel and water depth slightly, to $\beta=3\upi/4$ and $H=0.21$ at the same source velocity (same $\Fr$), creates the opposite situation; now the waves go from supercritical to subcritical as the shear is increased through the same values. It is possible in the latter situation for the waves to become supercritical once more for even higher $\Frs$, but the required values of $\Frs$ for the second transition ($5.4$ in this example) are so large as to seem unrealistic in practice. 

The phenomenon may be understood in terms of dispersion of the waves propagating in the direction of ship motion, i.e., transverse waves, since criticality is the transverse waves becoming too slow to keep up with the source. The effect of finite depth is to limit the phase velocity to values $\leq \sqrt{gh}$ isotropically, while the sub-surface shear current will advance waves going downstream (as seen from the system where the surface is at rest) and retard upstream-propagating waves. In the bottom 6 panels of Fig.~\ref{fig:transition}, transverse waves which would otherwise be too slow to contribute to the stationary wake are helped along by the sub-surface current, rendering the situation sub-critical again when the shear is increased.

\section{Wave resistance}\label{sec:WR}

We present in this section, for the first time, numerical evaluation and discussion of the wave resistance on a moving pressure source in the presence of shear. We illustrate and discuss the interplay shear and finite depth affect the wave resistance. Calculations of the lateral wave ``resistance'', present for $\beta\neq0,\pi$, are reported here for the first time to our knowledge. 

\begin{figure}
  \centerline{\includegraphics[width=4in]{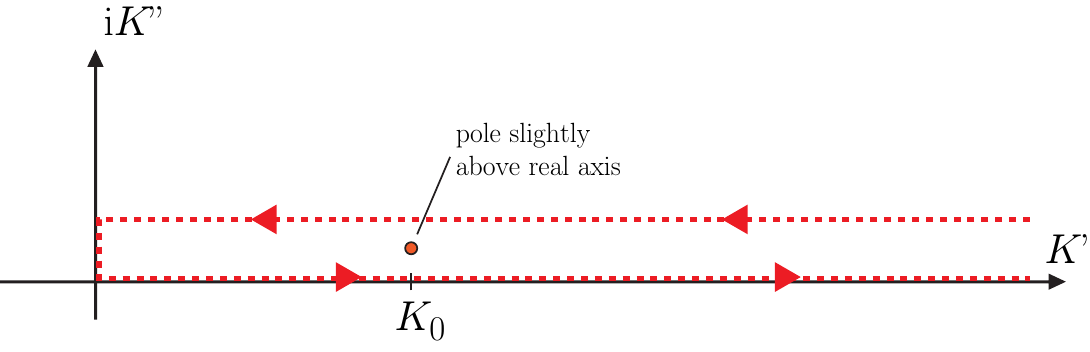}}
  \caption{Contour considered for the calculation of wave resistance.}
  \label{fig:Rcontour}
\end{figure}

In the same vain as Eq.~\eqref{zetanondim} we can write \eqref{Rgen} with pressure \eqref{pext} as
\bs
\begin{align}
  R =& \frac{bp_0^2}{4\upi^4\rho g \Fr^2}  \int_{-\upi}^{\upi}\frac{\rmd \gamma}{\cos\gamma} \lim_{\epsilon\to 0^+}J(\gamma) \\
  J(\gamma) =& \rmi\int_0^\infty \rmd K\frac{K^2\rme^{-2(K/2\upi)^2}\tanh KH }{\Gamma(K,\gamma)+\rmi \epsilon\Psi(K,\gamma)} \equiv \int_0^\infty\rmd K g(K).
\end{align}
\es
Just as for the integral \eqref{zetanondim}, provided a solution $K_0(\gamma)$ in Eq.~\eqref{K0def} exists there is a simple pole which, through the introduction of $\epsilon$, has been moved slightly off the real $K$ axis to a position which is slightly above the axis if $\cos\gamma<0$ and slightly below it if $\cos\gamma>0$ (see the discussion in Sec.~\ref{sec:nearfar}).

Consider now the case where the pole lies above the axis, and regard the closed contour in the complex $K$ plane shown in figure \ref{fig:Rcontour}. The contour consists of the real $K$ axis (the original integration path giving $J(\gamma)$) and a horizontal path parallel with the real axis but just far enough above to enclose the pole into the contour. While the quantity $R$ is real by construction, a finite value of $\epsilon$ introduces a small imaginary part which vanishes as $\epsilon\to 0$, so we treat $J(\gamma)$ as complex. Now note that because the imaginary unit $\rmi$ enters only as a prefactor and in front of $\epsilon$, the integral along the upper horizontal path (towards the left) equals $J^*(\gamma)$ (complex conjugate) plus a correction of order $\epsilon$. Hence we have by the Cauchy integral theorem that
\be
  \oint\rmd K g(K)  = 2\upi \rmi \res_{K=K_0}g(K) = J(\gamma) + J^*(\gamma) = 2\mathrm{Re} \{J(\gamma)\}\buildrel{\epsilon\to 0}\over{\longrightarrow}2J(\gamma).
\ee
If the pole is below the axis instead, the argument remains the same while closing the path in the lower halfplane, producing the opposite sign because the pole is now encircled in the negative sense, and we obtain all together
\begin{align}
  \frac{R}{R_0} =& \frac1{2\Fr^2}  \int_{-\upi}^{\upi}\frac{\rmd \gamma}{|\cos\gamma|}\frac{K_0^2\rme^{-K_0^2/2\upi^2}\tanh K_0H }{\Gamma'(K_0,\gamma)}\Theta[f_s(\gamma)-\Frh^2\cos^2\gamma]
\end{align}
where 
$R_0 = bp_0^2/(2\upi^3 \rho g)$, and the $\Theta$ function again ensures inclusion only of $\gamma$ for which $K_0(\gamma)$ exists [$f_s(\gamma)$  was defined in Eq.~\eqref{fs}]. For comparison with \citet{benzaquen14} we plot the quantity $R/R_0$ which corresponds to the function $f$ in their equation 18\footnote{Our result is a factor $2$ greater than that of \citet{benzaquen14} for comparable parameters; the latter reference seems to lose this factor somewhere between their equations 16 and 18.}. (The same result could be obtained by bypassing the pole on the real axis with a small semicircle, above or below as appropriate).

Hence the wave resistance is given by the pole at $K_0$ only, which is in fact physically obvious: only waves which satisfy the dispersion relation $K=K_0(\gamma)$ are allowed to propagate towards infinity and thereby remove energy from the source by way of wave resistance. 

\begin{figure}
  \includegraphics[width = \textwidth]{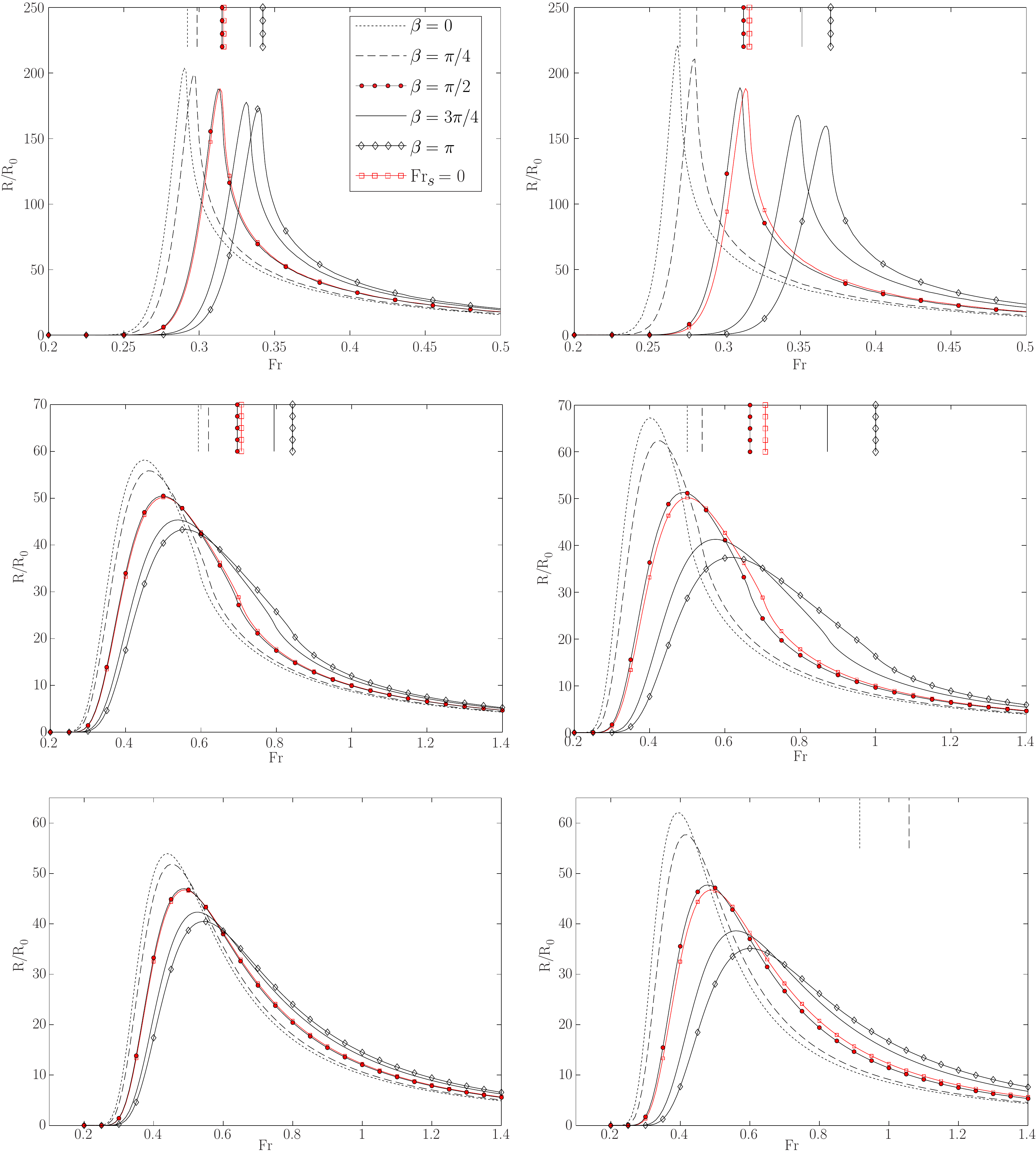}
  \caption{Wave resistance at depth $H=0.1$ (top row), $0.5$ (middle row) and $\infty$ (bottom row) for 
  $\Frsb=0.25$ (left column) and $0.5$ (right column). The vertical lines show the critical Froude numbers as given in Eq.~\eqref{FrC}.}
  \label{fig:WR}
\end{figure}

In his classical analysis nearly a century ago, \citet{havelock22} showed that for an axisymmetric pressure travelling in water of finite depth so that $H$ was of order $1$, the wave resistance showed a clear peak at a velocity just below the critical velocity $\sqrt{gh}$, and decreased rapidly for velocities higher than this. We observe the same trend at shallow and intermediate depth also in the presence of shear, while for $H\gtrsim 1$ the critical Froude number becomes of little consequence to the wave resistance. Wave resistance is calculated for three different depth ($H=0.1, 0.5$ and $\infty$) and shown in Fig.~\ref{fig:WR} for different directions of motion and increasing values of shear, parameterised through $\Frsb$ defined in Eq.~\eqref{Frsb}.

The importance of the critical velocity to wave resistance in shallow water, but not in deep water, can be understood from considerations of dispersion. In deep water there is strong dispersion which causes a gradual transition from a wake dominated by transverse waves propagating along the direction of ship motion, to diverging waves propagating almost normal to the direction of motion. Since wave resistance equals the rate at which forward momentum is imparted to the waves, transverse waves must contribute more to wave resistance than a diverging wave of the same absolute momentum but directed almost normal to the ship's motion. For this reason deep water wave resistance naturally peaks at a value of $\Fr$ corresponding to the increasing importance of diverging waves. If the critical velocity occurs for a value of $\Fr$ where transverse waves are still strongly present, however, wave resistance experiences a sudden drop near the critical value because transverse waves vanish. This occurs in shallow water, shown in the upper left panel of Fig.~\eqref{fig:WR}, and could also occur for very strong shear.

In general the effect of shear upon wave resistance is twofold: to shift the velocity of maximum resitance, and to modify the value of the maximum resistance. For motion against the direction of shear ($\beta=0$) the peak resistance is higher and is found at smaller values of $\Fr$ compared to zero shear, while for the source moving with the shear ($\beta=\upi$) the peak is lower and shifted to higher $\Fr$. In all cases the wave resistance for side-on shear ($\beta=\upi/2$) is very close to that found without shear current. All shear  effects are stronger for higher values of $\Frsb$ as can be expected. The effect of shear is most dramatic in shallow water, where wave resistance has a sharp peak near the critical Froude number. For vessels operating in the shallows at Froude numbers near the critical, inclusion of the effect of shear seems to be crucial for a realistic calculation of wave resistance. 

Firstly, the velocity at which wave resistance peaks, is lower for upstream ship motion. In fact, inspection reveals that also other effects of increasing velocity, such as the transition from transverse to diverging wave dominated patterns, occur at lower $\Fr$ for upstream ship motion ($\beta\sim 0$) than for downstream ($\beta\sim\pi$). Because waves with an upstream propagation component are retarded by the sub-surface shear flow, an upstream-going ship has, in a rough sense, a higher effective velocity as far as the waves are concerned, and effects of increasing velocity are thus shifted to lower values of $\Fr$. The opposite is the case for downstream motion. 

To understand why wave resistance has a higher peak for upstream motion than for downstream, it may be most instructive to consider the directional dependence of group velocity in the presence of shear, discussed in detail by \cite{ellingsen14b}. First, notice that wave resistance peaks at a value of $\Fr$ where the waves are dominated by transverse waves, i.e., waves propagating approximately in direction $\beta$. Ship waves have phase velocity which equals the ship velocity along the line of motion, and since gravity waves have a group velocity smaller than their phase velocity, the waves, after being generated (at the bow, say) are left behind as the boat moves forward. A wave can contribute to wave resistance only while it remains in the near field: once left behind its influence is no longer felt by the ship. As discussed by \cite{ellingsen14b} (and illustrated in the context of ring waves), waves travelling upstream have weakened dispersion, and the shear makes the group velocity approach the phase velocity. The time the wave spends in the ship's near zone is proportional to the difference between group and phase velocity, hence the wave travelling upstream will remain in the ship's near-zone for a longer time, increasing the effect on wave resistance. For the same reason, the effects of transient waves created by the source from maneuvering, say, can continue to affect the wave resistance for a long time for upstream motion, but are quickly whisked away for downstream motion \citep{li15b}.

For all wave resistance calculations it must be noted that wave resistance depends on the shape of the wave source, and the calculations herein, performed for a circular source, is only an illustration and rough indication of the effect of shear upon the wave resistance on a particular hull. The methodology can however be applied to more realistic shapes, at the cost of introducing further parameters.

\subsection{Lateral wave ``resistance''}

\begin{figure}
  \includegraphics[width = \textwidth]{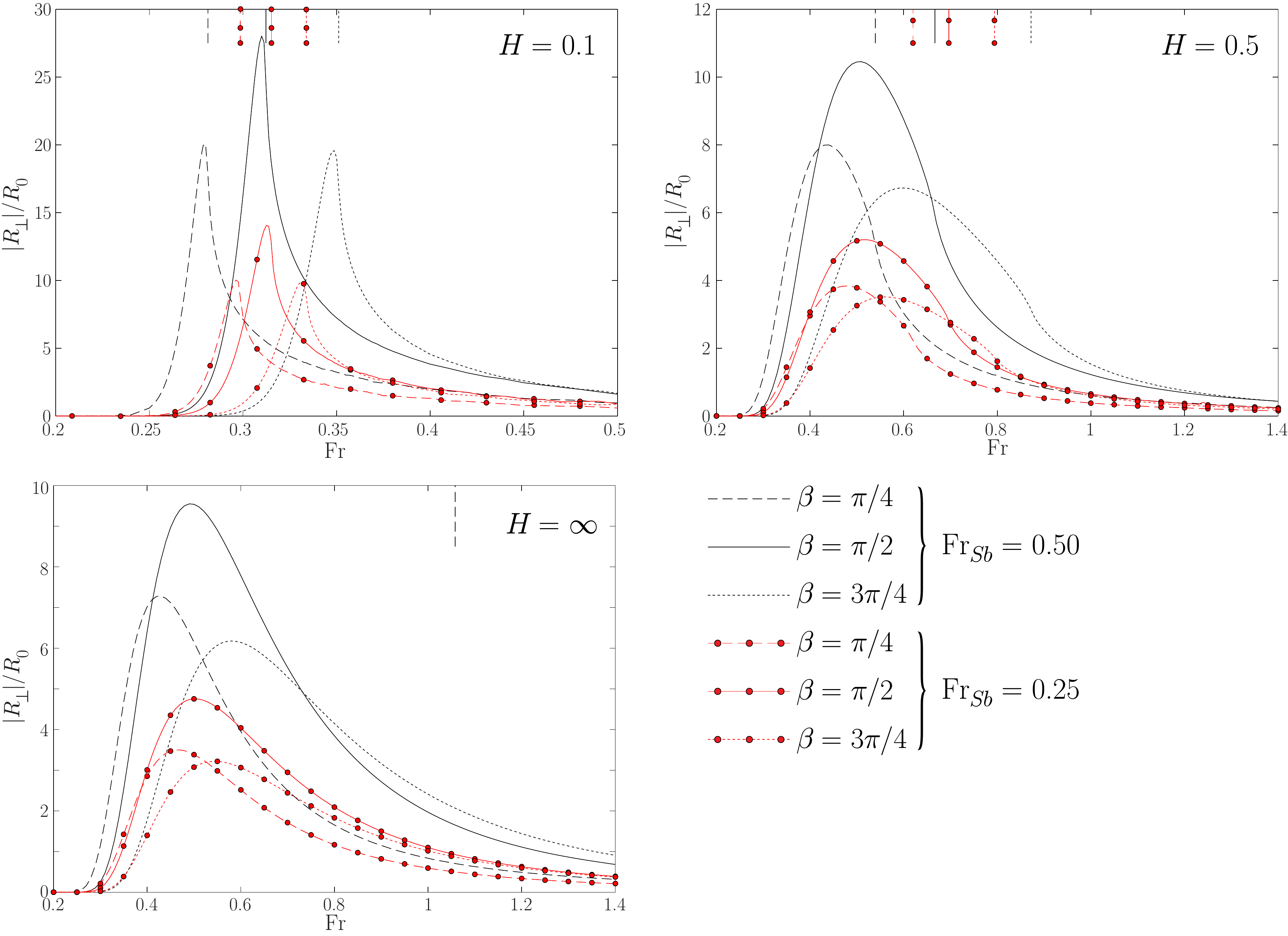}
  \caption{Lateral ``wave resistance'' for the same values of $H,\beta$ and $\Frsb$ as in figure \ref{fig:WR} (for $\beta=0,\upi$ the lateral force is zero). The vertical lines show the critical Froude numbers as given in Eq.~\eqref{FrC}. The absolute value of $R_\perp$ is shown; for $0\leq \beta\leq \upi$, $R_\perp\leq 0$.}
  \label{fig:WRperp}
\end{figure}

Now consider the lateral ``wave resistance'' $R_\perp$ which is calculated exactly like $R$ was above, while noting that $(\mathbf{e}_z\times\bV)\bcdot\bk = kV\sin\gamma$,
\be
   \frac{R_\perp}{R_0} = \frac{1}{2\Fr^2}  \int_{-\upi}^{\upi}\rmd \gamma\frac{\tan\gamma}{|\cos\gamma|}\frac{K_0^2\rme^{-K_0^2/2\upi^2}\tanh K_0H }{\Gamma'(K_0,\gamma)}\Theta[f_s(\gamma)-\Frh^2\cos^2\gamma].
\ee
The lateral force is plotted in Fig.~\ref{fig:WRperp} for the same values of $H$ as used in Fig.~\ref{fig:WR}. Just like the standard wave resistance, the lateral component also tends to pull the source in the direction where the wave field is strongest, i.e., downstream (as seen from the system where the undisturbed water surface is at rest). The lateral force behaves more or less like the wave resistance as a function of parameters, but is smaller in magnitude.  Its maximum value is when the current is approximately side-on ($\beta\sim\upi/2$), in which case its value can amount to about $10$-$20\%$ of the value of the corresponding no-shear wave resistance when the shear is strong ($\Frsb\sim 1$).

\section{Further discussion and concluding remarks}

We have presented a comprehensive theory for linear ship waves on a shear current of uniform vorticity, including the effects of finite depth. A finite water depth and nonzero vorticity each affect the dispersion of the water waves, and the resulting pattern is governed by the two effects in subtle combination. In particular, the phenomenon of a critical velocity at which the wake becomes very broad and above which transverse waves vanish, can be caused by either finite depth or nonzero shear, and in the presence of both the behaviour is intricate. 

We derive an explicit formula for the critical velocity for our system as a function of two different Froude numbers (with respect to water depth, and an intrinsic Froude number) and the angle between current and the source's line of motion. The phenomenon may be fully understood in terms of dispersion of the waves propagating in the direction of ship motion, i.e., transverse waves, since criticality is the transverse waves becoming too slow to keep up with the source. The effect of finite depth is to limit the phase velocity to values $\leq \sqrt{gh}$ isotropically, while the sub-surface shear current will advance waves going downstream (as seen from the system where the surface is at rest) and retard upstream-propagating waves. While it was previously shown that increasing the shear strength in deep water could cause transition to sub- to supercritical waves \citep{ellingsen14a}, in the presence of finite depth, increasing shear strength can also cause the opposite transition for mainly downstream ship motion.

It is a general observation that the presence of a sub-surface shear current in a system where the surface is at rest will tend to alter the ``effective velocity" of the ship (in a rough sense) relative to the water as far as the waves are concerned. While this notion provides a rough  prediction of some qualitative effects of shear upon various aspects of ship waves in shear conditions, it cannot capture the full picture since the dispersion properties of waves in the presence of shear have a subtle directional dependence. 

A source generating ship waves will feel a resistance force because waves carry energy away from the source. In the presence of a shear current there will in general also be a lateral component to this force, because the wake is asymmetrical about the ship's line of motion unless its angle with the shear current is exactly $0$ or $\upi$. We derive formulae for both the normal and lateral wave resistance force and analyse its dependence on the source velocity (or Froude number $\Fr$) for different amounts of shear and different directions of motion. For a circular source the well known dependence of wave resistance $R$ on Froude number is observed --- $R$ rises to a maximum value before falling off again at higher $\Fr$ --- and the role of the shear current is twofold. Firstly, the velocity of maximum wave resistance, $\Fr_\text{max}$, is lowered for directions moving the source up against the shear, but increased when the motion has a component downstream with the shear (as seen from a system at which the surface is at rest). Secondly the maximum wave resistance is increased for directions against the shear, and decreased for downstream directions.

The lateral ``wave resistance'' behaves much the same way as a function of Froude number, it is maximal when the current is approximately normal to the direction of motion, and tends to zero for directions directly upstream or downstream.

\appendix

\section{Details of path integration}\label{app:farfield}

In this appendix may be found details of the calculation of path integral $I(\gamma)$ from Eq.~\eqref{zetanondim}

\subsection{Contribution from the pole}\label{app:polecontrib}
The position of the pole in the integrand of $I(\gamma)$ is now at 
$
  K_\text{pole} = K_0 - \rmi\epsilon F(K_0,\gamma)
$
where 
\be\label{F}
  F(K_0,\gamma)
  = \frac{\Psi(K_0,\gamma)}{\Gamma'(K_0,\gamma)} 
  = \frac{2K_0\Fr^2\cos\gamma + \Frs\cos(\gamma+\beta)\tanh(K_0H)}{\Fr^2\cos^2\gamma - f_s(\gamma)H\sech^2( K_0H)}
\ee
where $\Gamma'(K,\gamma) = \partial \Gamma(K,\gamma)/\partial K$. The key property of $F(K_0,\gamma)$ is its \emph{sign}, which determines which side of the real $K$ axis the pole lies. To wit, we note that in order for the pole to lie inside the contour $\Lambda$, $F$ and $\cos(\gamma-\phi_\beta)$ must have opposite signs. It is also necessary that $K_0(\gamma)>0$ as discussed above. If we assume that for some $\gamma$ a solution $K_0(\gamma)>0$ exists, we can use definition Eq.~\eqref{K0def} to write
\be
  F(K_0,\gamma) = \frac{f_s(\gamma)+1}{\Fr^2\cos^3\gamma}\frac{2\sinh^2 \varkappa}{\sinh 2\varkappa - 2\varkappa}
\ee
with $\varkappa=K_0(\gamma)H$. It follows from the criterion Eq.~\eqref{critGamma} that for $K_0(\gamma)>0$ to exist it is necessary that $f_s(\gamma)>0$. Moreover one easily ascertains that $2\sinh^2 \varkappa/(\sinh 2\varkappa - 2\varkappa)$ is a positive function of $\varkappa$ for all $\varkappa>0$. In other words, 
$
  \sg[F(K_0,\gamma)] = \sg(\cos\gamma),
$
provided $K_0(\gamma)>0$ ($\sg$ is the signum function). Hence the pole lies inside the closed contour provided $\cos\gamma$ and $\cos(\gamma-\phi_\beta)$ have opposite signs, i.e., $\cos\gamma \cos(\gamma-\phi_\beta)<0$.

Hence we may write using the Cauchy integral theorem, noting that the pole is encircled in the positive sense if $\cos(\gamma-\phi_\beta)>0$ and \emph{vice versa}, 
\begin{align*}
  \oint_\Lambda \rmd K f(K,\gamma) =& 2\upi\rmi \Theta(K_0)\Theta[-\cos(\gamma-\phi_\beta)\cos\gamma]\sg[\cos(\gamma-\phi_\beta)]\Res_{K=K_\mathrm{pole}}\left\{f(K)\right\}\\
  =& I(\gamma) - \Isd (\gamma) - \Ic(\gamma)
\end{align*}
or, noting that $\sg[\cos(\gamma-\phi_\beta)]=-\sg[\cos\gamma]$ if the pole contributes,
\begin{align}\label{Ioint}
  I(\gamma)  =& \Isd (\gamma) + \Ic(\gamma)\notag \\
  &- 2\upi\rmi \Theta(K_0)\Theta[-\cos(\gamma-\phi_\beta)\cos\gamma]\sg[\cos\gamma]\Res_{K=K_\mathrm{pole}}\left\{f(K)\right\}
\end{align}
where the Heaviside functions $\Theta$ enforce the criteria for the pole being inside the integral, and the contributions from the parts of the path are
\begin{align}
  \Isd (\gamma) =&  \int_{|\Xi|}^\infty\rmd K f(K + \rmi \Xi,\gamma),\\
  \Ic(\gamma) =& (1\pm\rmi) \int_0^{|\Xi|}\rmd K f[(1\pm \rmi) K ,\gamma]\label{Ic}.
\end{align}
We use the shorthand
$  \Xi = 2\upi^2 X\cos(\gamma-\phi_\beta),$
whereby $K_\text{s.d.}= K+\rmi\Xi$. The sign to be taken in Eq.~\eqref{Ic} is the sign of $\Xi$.
Thus, by Eq.~\eqref{Ioint} the integral $I(\gamma)$ can be written in terms of a steepest descent integral, a connection path, and possibly the residue of the pole. 

\subsection{Asymptotic falloff of path integrals}\label{app:nearfield}

Consider now large distances $X$ to consider asymptotic behaviour. We presume therefore in the following that $|\Xi|\to \infty$.

Inserting $K_\text{s.d.}$, the steepest descent integral becomes ($\epsilon$ is only of interest inasmuch as it moves the pole off the real $K$ axis, and can be set to zero in the following)
\be\label{Isdint}
  \Isd = \rme^{-\Xi^2/2}\int_{|\Xi|}^\infty \rmd K\frac{\Ksd \rme^{-(K/2\upi)^2}\tanh(\Ksd H)}{\Gamma(\Ksd,\gamma)}.
\ee
Numerical inspection reveals that the integral in \eqref{Isdint} is a nearly periodic function of $\Xi$ with sharp peaks at $\Xi H = (n+\half)\upi$ ($n=1,2,3,...$) and which is everywhere of order unity. The prefactor $\exp(-\Xi^2/2)$ and the fact that the integral starts at the (already very large) lower limit $|\Xi|$ ensure that the integral is exponentially small as $\Xi\to \infty$.

Next considering $\Ic$, we write
\be
  \Ic = (1\pm\rmi)^2\int_0^{|\Xi|}\rmd K \frac{K\rme^{-2\rmi K^2/(2\upi)^2}\tanh(1\pm \rmi)KH}{\Gamma[(1\pm\rmi)K]}\rme^{(\rmi\Xi - |\Xi|)K/2\upi^2}.
\ee
Asymptotic analysis of this integal in itself is straightforward, by noting that its main contribution comes from small values of $K$ of order $1/\Xi$. However, the resulting expression yields a non-integrable function of $\gamma$, so an asymptotic analysis of the full double integral is necessary. We satisfy ourselves by performing the integral numerically and find that the contribution to $\zeta$ from the connection integral falls off slightly faster than $X^{-1}$ as $X\to\infty$ (most likely a logarithmic contribution is involved). The surface deformation in the far field falls off as $X^{-1/2}$ \citep[e.g.][]{ellingsen14a} as is required by energy conservation, hence the connection integral is part of the near field. 

We have shown that $\Isd$ and $\Ic$ fall off faster than the far field solution when $X\to \infty$. More detailed asymptotic analysis of $\Isd$ and $\Ic$ is of course possible, yet given the complexity of the integrals involved and the fact that our primary interest is the far field, we shall be content here with this brief treatment.

\section{Note on the use of the Sokhotski--Plemelj theorem}\label{app:SP}

\begin{figure}
  \includegraphics[width=\textwidth]{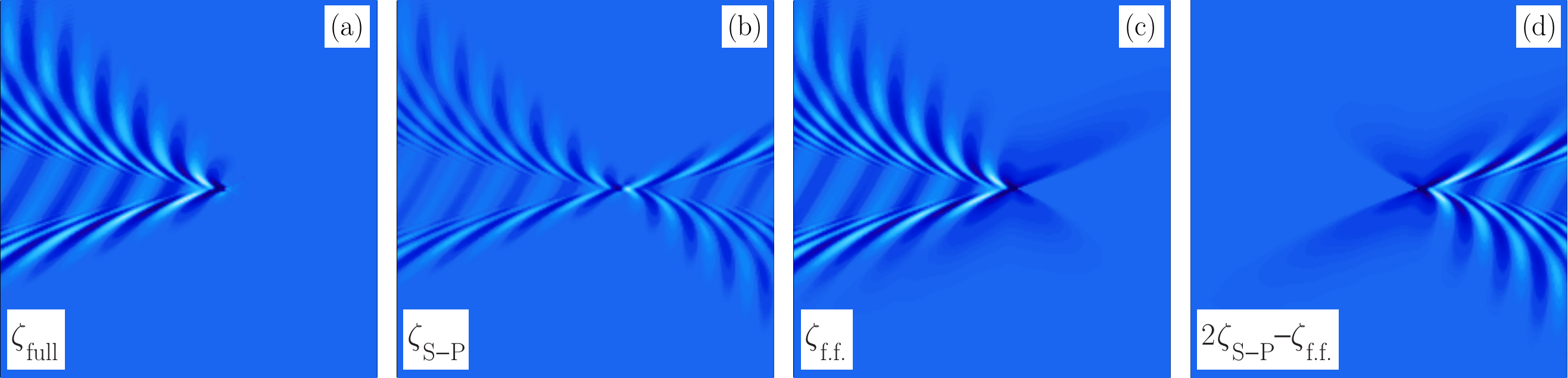}
  \caption{Comparison of wave fields $\zeta(\bx)$ calculated using (a) full expression from Eq.~\eqref{zetaGen} with a small, nonzero value of $\epsilon$, (b) the Sokhotksy-Plemelj expression $\zeta_\text{S-P}$ (with integration range $-\upi<\gamma<\upi$), (c) our far-field expression $\zff$ from Eq.~\eqref{zetafar2}, and the combination $2\zeta_\text{S-P}-\zff$. Parameters are $\Fr=\Frs=0.8$, $\Frh=0$, and $\beta=\upi/2$ in all panels. }
  \label{fig:SP}
\end{figure}

In the literature, the so-called Sokhotski--Plemelj theorem has sometimes been used in order to extract the far-field contribution to the wave field \citep{raphael96,darmon14,ellingsen14a,benzaquen14}. We show here that while the correct far-field can be obtained this way in parts of the plane (modulo a prefactor 2 for the wave amplitude), what is obtained is in fact a combination of the far-field allowed by the radiation condition and the spurious ``far-field'' which is physically illegal because it corresponds to waves travelling from future to past (or alternatively: originate at infinity and converge at the source).

Let us consider the surface deformation $\zeta$ and consider the case of deep water for simplicity (the principle is the same for finite water depth), in which case $\zeta$ from Eq.~\eqref{zetaGen} can be written on the general form 
\be\label{zSP}
  \zeta = \zeta_0 \pipiint \rmd \gamma \lim_{\epsilon\to 0}\int_0^\infty  \frac{\rmd k\,f(k)}{k-k_0-\rmi\epsilon\phi}
\ee
where $\zeta_0$ is a constant and $\phi$ is a function of $(k,\gamma)$ which can take either sign. We now wish to evaluate just the far-field contribution to the  $k$-integral. The procedure of, e.g., \citet{darmon14} is now to use the so-called Sokhotski--Plemelj (SP) theorem to evaluate the contribution from the simple pole at $k=k_0$, which is then identified as the far-field. 

The SP theorem is very simple to derive and results from simply multiplying the integrand by $k-k_0 + i\epsilon\phi$ in numerator and denominator and splitting into two terms:
\begin{align}
  &\lim_{\epsilon\to0}\int_0^\infty\rmd k\frac{f(k)}{k-k_0-\rmi\epsilon\phi} \notag\\
  &= \rmi\upi\sg(\phi)\lim_{|\phi|\epsilon\to 0}\int_0^\infty\rmd k\frac{\epsilon|\phi|f(k)}{\upi[(k-k_0)^2+\epsilon^2\phi^2]}+\lim_{\epsilon\to 0}\int_0^\infty\rmd k\frac{(k-k_0)^2}{(k-k_0)^2+\epsilon^2\phi^2}\frac{f(k)}{k-k_0} \notag\\
  &= \rmi\upi\sg(\phi_0)f(k_0) + \mathcal{P}\int_0^\infty \rmd k \frac{f(k)}{k-k_0}\label{appb}
\end{align}
where $\mathcal{P}$ denotes the principal value, and $\phi_0=\phi(k_0,\gamma)$. The last form is obtained by noting that $\varepsilon/\upi(x^2+\varepsilon^2)\to \delta(x)$ as $\varepsilon\to0^+$, and recognising a definition of the principal value. The theorem is of course valid, but the potential danger lies in now identifying the first term in the last form of \eqref{appb}, proportional to $f(k_0)$ as the far-field, and the second term as the near field. 

We can see this most easily by regarding the term which becomes the $\delta$-function, purportedly the far-field contribution according to the SP procedure. In a slight change of notation from the above this term alone can be written
\be\label{zSP2}
  \lim_{\epsilon\to 0}\int_0^\infty\rmd k\frac{i\epsilon \phi f(k)}{(k-k_0)^2+\epsilon^2\phi^2} = \lim_{\epsilon\to 0}\int_0^\infty\rmd k\frac12\left[\frac{f(k)}{k-k_0-\rmi\epsilon\phi}-\frac{f(k)}{k-k_0+\rmi\epsilon\phi} \right]
\ee
where we have expanded in partial fractions. Now comparing with the original, full expression \eqref{zSP} we see that the porported ``far-field" solution is in fact half the full wave field minus half of the wave field obtained from swapping the sign of $\epsilon$, i.e., exactly the waves which the radiation condition is supposed to exclude. 

Apart from the factor $1/2$ the situation is not too serious, because the real wake in equation \eqref{zSP} lies behind the ship, whereas the spurious ``wake from the future" in the second term of \eqref{zSP2} lies in front of it. The situation is illustrated in Fig.~\ref{fig:SP} Indeed the ``far-field'' wave expressions in \citet{darmon14}, \citet{ellingsen14a} give waves also antisymmetrically in front of the ship which must be manually removed (for example by simply not plotting them). The procedure employed in our Section \ref{sec:nearfar}, while somewhat more elaborate, avoids this problem.

\bibliographystyle{jfm}
\bibliography{shipwave}

\begin{thebibliography}{35}
\expandafter\ifx\csname natexlab\endcsname\relax\def\natexlab#1{#1}\fi

\bibitem[Bender \& Orszag(1991)]{bender91}
{\sc Bender, C.~M. \& Orszag, S.~A.} 1991 {\em Advanced Mathematical Methods
  for Scientists and Engineers\/}. Springer.

\bibitem[Benzaquen {\em et~al.\/}(2014)Benzaquen, Darmon \&
  Rapha{\"e}l]{benzaquen14}
{\sc Benzaquen, M., Darmon, A. \& Rapha{\"e}l, E.} 2014 Wake pattern and wave
  resistance for anisotropic moving disturbances. {\em Phys. Fluids\/} {\bf
  26}, 092106.

\bibitem[Brown {\em et~al.\/}(1989)Brown, Buchsbaum, Hall, Penhune, Schmitt,
  Watson \& Wyatt]{brown89}
{\sc Brown, E.~D., Buchsbaum, S.~B., Hall, R.~E., Penhune, J.~P., Schmitt,
  K.~F., Watson, K.~M. \& Wyatt, D.~C.} 1989 Observations of a nonlinear
  solitary wave packet in the {K}elvin wake of a ship. {\em J. Fluid Mech.\/}
  {\bf 204}, 263--293.

\bibitem[B\"{u}hler(2009)]{buhler09}
{\sc B\"{u}hler, O.} 2009 {\em Waves and Mean Flow\/}. Cambridge University
  Press.

\bibitem[Craik(1968)]{craik68}
{\sc Craik, A. D.~D.} 1968 Resonant gravity--wave interactions in a shear flow.
  {\em J.~Fluid Mech.\/} {\bf 34}, 531--549.

\bibitem[Darmon {\em et~al.\/}(2014)Darmon, Benzaquen \& Rapha\"{e}l]{darmon14}
{\sc Darmon, A., Benzaquen, M. \& Rapha\"{e}l, E.} 2014 Kelvin wake pattern at
  large {F}roude numbers. {\em J.~Fluid Mech.\/} {\bf 738}, R3.

\bibitem[Ellingsen(2014{\natexlab{{\em a\/}}})]{ellingsen14b}
{\sc Ellingsen, S.~{\AA}} 2014{\natexlab{{\em a\/}}} Initial surface
  disturbance on a shear current: The {C}auchy-{P}oisson problem with a twist.
  {\em Phys. Fluids\/} {\bf 26}, 082104.

\bibitem[Ellingsen(2014{\natexlab{{\em b\/}}})]{ellingsen14a}
{\sc Ellingsen, S.~\AA.} 2014{\natexlab{{\em b\/}}} Ship waves in the presence
  of uniform vorticity. {\em J. Fluid Mech.\/} {\bf 742}, R2.

\bibitem[Ellingsen(2016)]{ellingsen16}
{\sc Ellingsen, S.~{\AA}.} 2016 Oblique waves on a vertically sheared current
  are rotational. {\em Eur. J. Mech.-B/Fluids\/} {\bf 56}, 156--160.

\bibitem[Ellingsen \& Brevik(2014)]{ellingsen14c}
{\sc Ellingsen, S.~\AA. \& Brevik, I} 2014 How linear surface waves are
  affected by a current with constant vorticity. {\em Eur.\ J.~Phys.\/} {\bf
  35}, 025005.

\bibitem[Ellingsen \& Tyvand(2016)]{ellingsen15b}
{\sc Ellingsen, S.~{\AA} \& Tyvand, P.~A.} 2015 Oscillatory point source in
  flow of uniform shear in three dimensions. {\em J. Fluid Mech.\/} (in press).

\bibitem[Faltinsen(2005)]{faltinsen05}
{\sc Faltinsen, O.~M.} 2005 {\em Hydrodynamics of High--Speed Marine
  Vehicles\/}. Cambridge University Press.

\bibitem[Havelock(1908)]{havelock08}
{\sc Havelock, T.~H.} 1908 The propagation of groups of waves in dispersive
  media, with application to waves on water produced by a travelling
  disturbance. {\em Proc.\ R.~Soc.\ London A\/} {\bf 81}, 398--430.

\bibitem[Havelock(1917)]{havelock17}
{\sc Havelock, T.~H.} 1917 Some cases of wave motion due to a submerged
  obstacle. {\em Proc. R. Soc. London A\/} {\bf 100}, 520--532.

\bibitem[Havelock(1919)]{havelock19}
{\sc Havelock, T.~H.} 1919 Wave resistance: Some cases of three-dimensional
  fluid motion. {\em Proc.\ R.~Soc.\ London A\/} {\bf 95}, 354--365.

\bibitem[Havelock(1922)]{havelock22}
{\sc Havelock, T.~H.} 1922 The effect of shallow water on wave resistance. {\em
  Proc.\ R.~Soc.\ London A\/} pp. 499--505.

\bibitem[He {\em et~al.\/}(2015)He, Zhang, Zhu, Wu, Yang, Noblesse, Gu \&
  Li]{he15}
{\sc He, J., Zhang, C., Zhu, Y., Wu, H., Yang, C.-J., Noblesse, F., Gu, X. \&
  Li, W.} 2015 Comparison of three simple models of kelvin's ship wake. {\em
  Eur. J. Mech.-B/Fluids\/} {\bf 49}, 12--19.

\bibitem[Johnson(1990)]{johnson90}
{\sc Johnson, R.~S.} 1990 Ring waves on the surface of shear flows: a linear
  and nonlinear theory. {\em J. Fluid Mech.\/} {\bf 215}, 145--160.

\bibitem[Li \& Ellingsen(2015)]{li15}
{\sc Li, Y. \& Ellingsen, S.~\AA.} 2015{\natexlab{{\em a\/}}} Initial value
  problems for water waves in the presence of a shear current. In {\em
  Proceedings of the 25th International Offshore and Polar Engineering
  Conference (ISOPE)\/}, pp. 543--549.

\bibitem[Li \& Ellingsen(2016)]{li15b}
{\sc Li, Y. \& Ellingsen, S.~\AA.} 2015{\natexlab{{\em b\/}}} Water waves from
  general, time-dependent surface pressure distribution in the presence of a
  shear current. {\em Intl J. Offshore Polar Engng\/} (in press).

\bibitem[Lighthill(1978)]{lighthill78}
{\sc Lighthill, J.} 1978 {\em Waves in Fluids\/}. Cambridge University Press.

\bibitem[Mc{H}ugh(1994)]{mchugh94}
{\sc Mc{H}ugh, J.~P.} 1994 Surface waves on an inviscid shear flow in a
  channel. {\em Wave Motion\/} {\bf 19}, 135--144.

\bibitem[Moisy \& Rabaud(2014)]{moisy14}
{\sc Moisy, F. \& Rabaud, M.} 2014 Mach-like capillary-gravity waves. {\em
  Phys. Rev. E\/} {\bf 90}, 023009.

\bibitem[Munk {\em et~al.\/}(1987)Munk, Scully-Power \& Zachariasen]{munk87}
{\sc Munk, W.~H., Scully-Power, P. \& Zachariasen, F.} 1987 The bakerian
  lecture, 1986. ships from space {\em Proc. R. Soc. London A\/} {\bf 412}~(1843), 231--254.

\bibitem[Noblesse {\em et~al.\/}(2014)Noblesse, He, Zhu, Hong, Zhang, Zhu \&
  Yang]{noblesse14}
{\sc Noblesse, F., He, J., Zhu, Y., Hong, L., Zhang, C., Zhu, R. \& Yang, C.}
  2014 Why can ship wakes appear narrower than kelvin's angle? {\em Eur. J.
  Mech.-B/Fluids\/} {\bf 46}, 164--171.

\bibitem[Peregrine(1976)]{peregrine76}
{\sc Peregrine, D.~H.} 1976 Interaction of water waves and currents. {\em Adv.
  Appl. Mech.\/} {\bf 16}, 9--117.

\bibitem[Pethiyagoda {\em et~al.\/}(2014)Pethiyagoda, McCue \&
  Moroney]{pethiyagoda14}
{\sc Pethiyagoda, R, McCue, S.~W. \& Moroney, T.~J.} 2014 What is the apparent
  angle of a kelvin ship wave pattern? {\em J. Fluid Mech.\/} {\bf 758},
  468--485.

\bibitem[Pethiyagoda {\em et~al.\/}(2015)Pethiyagoda, McCue \&
  Moroney]{pethiyagoda15}
{\sc Pethiyagoda, R, McCue, S.~W. \& Moroney, T.~J.} 2015 Wake angle for
  surface gravity waves on a finite depth fluid. {\em Phys. Fluids\/} {\bf 27},
  061701.

\bibitem[Rabaud \& Moisy(2013)]{rabaud13}
{\sc Rabaud, M. \& Moisy, F.} 2013 Ship waves: Kelvin or mach angle? {\em
  Phys.~Rev.\ Lett.\/} {\bf 110}, 214503.

\bibitem[Rapha\"{e}l \& de~Gennes(1996)]{raphael96}
{\sc Rapha\"{e}l, E. \& de~Gennes, P.-G.} 1996 Capillary gravity waves caused
  by a moving disturbance: wave resistance. {\em Phys.\ Rev.\ E\/} {\bf 53},
  3448--3455.

\bibitem[Reed \& Milgram(2002)]{reed02}
{\sc Reed, A.~M. \& Milgram, J.~H.} 2002 Ship wakes and their radar images.
  {\em Annu. Rev. Fluid Mech.\/} {\bf 34}, 469--502.

\bibitem[Thomson(1887)]{kelvin1887}
{\sc Thomson, Sir~W.} 1887 On ship waves. {\em Proc.\ Inst.\ Mech.\ Eng.\/}
  {\bf 38}, 409--434.

\bibitem[Wehausen(1973)]{wehausen73}
{\sc Wehausen, J.~W.} 1973 The wave resistance of ships. {\em Adv. Appl.
  Mech.\/} {\bf 13}, 93--245.

\bibitem[Zhang {\em et~al.\/}(2015)Zhang, He, Zhu, Yang, Li, Zhu, Lin \&
  Noblesse]{zhang15}
{\sc Zhang, C., He, J., Zhu, Y., Yang, C.-J., Li, W., Zhu, Y., Lin, M. \&
  Noblesse, F.} 2015 Interference effects on the kelvin wake of a monohull ship
  represented via a continuous distribution of sources. {\em Eur. J.
  Mech.-B/Fluids\/} {\bf 51}, 27--36.

\bibitem[Zhu {\em et~al.\/}(2015)Zhu, He, Zhang, Wu, Wan, Zhu \&
  Noblesse]{zhu15}
{\sc Zhu, Y., He, J., Zhang, C., Wu, H., Wan, D., Zhu, R. \& Noblesse, F.} 2015
  Farfield waves created by a monohull ship in shallow water. {\em Eur. J.
  Mech.-B/Fluids\/} {\bf 49}, 226--234.

\end{thebibliography}

\end{document}